\documentclass{article}

\usepackage{cite}
\usepackage[linesnumbered,ruled,vlined]{algorithm2e}
\usepackage{amsmath,amsthm,mathtools,bm,amsfonts,amssymb,array}

\usepackage{arxiv}

\usepackage[utf8]{inputenc} 
\usepackage[T1]{fontenc}    
\usepackage{hyperref}       
\usepackage{url}            
\usepackage{booktabs}       
\usepackage{amsfonts}       
\usepackage{nicefrac}       
\usepackage{microtype}      
\usepackage{lipsum}		
\usepackage[pdftex]{graphicx}
\usepackage{doi}

\usepackage{tikz}
\usepackage{textcomp}
\newcommand\copyrighttext{%
\footnotesize \textcopyright 2020 IEEE. Personal use of this material is permitted.
Permission from IEEE must be obtained for all other uses, in any current or future
media, including reprinting/republishing this material for advertising or promotional
purposes, creating new collective works, for resale or redistribution to servers or
lists, or reuse of any copyrighted component of this work in other works.

\doi{10.1109/TMM.2021.3079695}}
\newcommand\copyrightnotice{%
\begin{tikzpicture}[remember picture,overlay]
\node[anchor=south,yshift=20pt] at (current page.south) {\fbox{\parbox{\dimexpr\textwidth-\fboxsep-\fboxrule\relax}{\copyrighttext}}};
\end{tikzpicture}%
}

\newcommand\blfootnote[1]{%
  \begingroup
  \renewcommand\thefootnote{}\footnote{#1}%
  \addtocounter{footnote}{-1}%
  \endgroup
}

\title{An FCNN-Based Super-Resolution mmWave Radar Framework for Contactless Musical Instrument Interface}

\author{ \href{https://orcid.org/0000-0002-3388-4805}{\includegraphics[scale=0.06]{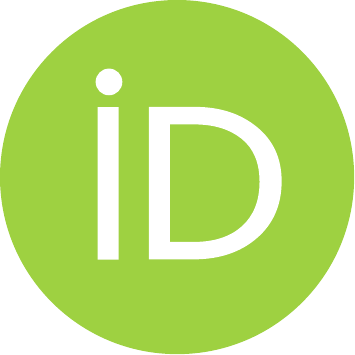}\hspace{1mm}Josiah W. Smith} \\
	Department of Electrical and Computer Engineering\\
	The University of Texas at Dallas\\
	Richardson, TX 75080 \\
	\texttt{josiah.smith@utdallas.edu} \\
	\And
	\href{https://orcid.org/0000-0003-0455-7597}{\includegraphics[scale=0.06]{orcid.pdf}\hspace{1mm}Orges Furxhi} \\
	Camera Systems and Computational Imaging\\
	IMEC-USA\\
	Kissimee, FL 34744 \\
	\texttt{orges.furxhi@imec-int.com} \\
	\And
	\href{https://orcid.org/0000-0001-7229-1765}{\includegraphics[scale=0.06]{orcid.pdf}\hspace{1mm}Murat Torlak} \\
	Department of Electrical and Computer Engineering\\
	The University of Texas at Dallas\\
	Richardson, TX 75080 \\
	\texttt{torlak@utdallas.edu} \\
}

\date{}

\renewcommand{\headeright}{\textit{IEEE Trans. Multimedia} (Accepted)}
\renewcommand{\undertitle}{\textit{IEEE Transactions on Multimedia} (Accepted)}
\renewcommand{\shorttitle}{An FCNN-Based Super-Resolution}

\hypersetup{
pdftitle={An FCNN-Based Super-Resolution mmWave Radar Framework for Contactless Musical Instrument Interface},
pdfsubject={eess.SP},
pdfauthor={Josiah W.~Smith, Orges Furxhi, Murat Torlak},
pdfkeywords={deep learning, human-computer interaction (HCI), fully-convolutional neural network (FCNN), millimeter-wave (mmWave), multiple-input multiple-output (MIMO), radar perception, super-resolution},
}

\begin{document}
\maketitle
\copyrightnotice

\begin{abstract}
In this article, we propose a framework for contactless human-computer interaction (HCI) using novel tracking techniques based on deep learning-based super-resolution and tracking algorithms. 
Our system offers unprecedented high-resolution tracking of hand position and motion characteristics by leveraging spatial and temporal features embedded in the reflected radar waveform. 
Rather than classifying samples from a predefined set of hand gestures, as common in existing work on deep learning with mmWave radar, our proposed imager employs a regressive full convolutional neural network (FCNN) approach to improve localization accuracy by spatial super-resolution. 
While the proposed techniques are suitable for a host of tracking applications, this article focuses on their application as a musical interface to demonstrate the robustness of the gesture sensing pipeline and deep learning signal processing chain. 
The user can control the instrument by varying the position and velocity of their hand above the vertically-facing sensor.
By employing a commercially available multiple-input-multiple-output (MIMO) radar rather than a traditional optical sensor, our framework demonstrates the efficacy of the mmWave sensing modality for fine motion tracking and offers an elegant solution to a host of HCI tasks.
Additionally, we provide a freely available software package and user interface for controlling the device, streaming the data to MATLAB in real-time, and increasing accessibility to the signal processing and device interface functionality utilized in this article. 
\end{abstract}

\blfootnote{This paper has supplementary downloadable material available at http://ieeexplore.ieee.org, provided by the authors. 
This material includes several video files demonstrating the proposed framework as a musical instrument and comparing its performance against existing techniques. 
The file "Music Example.mp4" is a video of our implementation of the proposed algorithms as a musical interface being played along with a backing track.
The files "Cross-Range Oscillation Tracking.mp4" and "Doppler Tracking.mp4" are videos of our FCNN-DPF tracking algorithms applied to track the cross-range oscillation rate and Doppler velocity, respectively. 
The file "Simple Tracking.mp4" is a video of the existing feature extraction and tracking techniques alongside the corresponding hand movements. 
The file "FCNN-DPF Tracking.mp4" is a video of our proposed FCNN-DPF feature extraction and particle filter tracking algorithms.
This material is $92$ MB in size.
}


\section{Introduction}
\label{sec:introduction}
Radar perception for human-computer interaction (HCI) on multiple-input-multiple-output (MIMO) millimeter-wave (mmWave) radars has emerged as a promising solution to a variety of sensing problems. 
The physical nature of millimeter-waves offers a safe method for high-resolution imaging where optical sensors may fail due to insufficient lighting, fog, or other line-of-sight interference. 
Additionally, mmWave sensors are considered less invasive than optical counterparts and promote user privacy.
Ultra-wideband MIMO devices enable centimeter-level spatial resolution with a small profile device.
As a result, precise spatial information of a target scene can be easily acquired from such imaging devices at a low cost.

Millimeter-wave sensors are relatively modern technology, but some of the earliest electronic interfaces were contactless devices for physically expressive musical control including the Radio Drum and Theremin \cite{winkler1995making}. 
Russian physicist Leon Theremin demonstrated his noncontact musical instrument in 1921, an interface controlled by the proximity of the musician's hand to an antenna using beat-frequency oscillators and a capacitive sensing apparatus \cite{skeldon1998physics}. 
More recently, computer vision approaches have been adopted for the innovation of contactless new musical interfaces (NMIs), most of which rely on optical camera solutions. 
Extensive prior work exists on optical-based NMIs using popular sensors such as the Microsoft Kinect and Leap Motion. 

In \cite{polfreman2011multi}, Polfreman uses the Kinect to track the 3-D position of both hands of a standing performer to construct a multi-modal instrument. 
Trail \textit{et al.} present a pitched percussion hyper-instrument to track the tips of two mallets simultaneously with the Kinect \cite{trail2012non}. 
Crossole, designed by Senturk \textit{et al.}, is a Kinect-based metainstrument that visualizes chord progressions as virtual blocks resembling a crossword puzzle \cite{senturk2012crossole}. 
Schramm \textit{et al.} use the Kinect to analyze and classify motions of an orchestral conductor \cite{schramm2015dynamic}. 

Alternatively, the popular Leap Motion controller is capable of modeling the entire hand, including the fingers, which allows for even more detailed hand posture-based gesture control to be explored for musical interface development. 
Using the Leap Motion sensor, Han \textit{et al.} developed two NMIs, \textit{Air Keys} and \textit{Air Pad}. 
\textit{Air Keys} tracks the motion and position of each finger to recognize when and which keys the musician is pressing and playing the desired notes. 
Similarly, \textit{Air Pad} tracks the hand position to create a 2-D virtual drum pad played by pressing specific regions in a 2-D horizontal plane, thus requiring accurate 3-D hand-tracking \cite{han2014lessons}. 
Hantrakul and Kaczmarek use the Leap Motion controller to track both hands for controlling MIDI (Musical Instrument Digital Interface) instruments and virtual effects \cite{hantrakul2014implementations}. 
Similarly, Leimu pairs the Leap Motion with an inertial measurement unit (IMU) demonstrating improved performance over the Leap Motion controller alone for musical interface \cite{brown2016leimu}. 
Other solutions have been attempted, such as employing non-invasive force sensing resistors to enhance ``traditional'' instruments by learning and monitoring for gestures performed by the musician \cite{tindale2011training}. 

In the optical HCI domain, \cite{nieto2013hand} proposes a musical interface using only a portable RGB camera to recognize hand gestures using a gesture classification technique. 
Akbari and Cheng developed a system to transcribe music played on a piano in real-time using optical cameras positioned to view the keys \cite{akbari2015real}. 
These projects have yielded high-performing real-time musical interfaces capable of consistent high-accuracy motion tracking but require several key design constraints, namely specific lighting conditions and line-of-sight. 
As shown in this article, mmWave sensors overcome these major obstacles while providing superior privacy through the means of advanced spatiotemporal algorithms.
However, little work has been done towards gestural musical interfaces on mmWave radar sensors using hand-tracking techniques. 
Even though extensive research exists on static and dynamic gesture recognition using deep learning models and mmWave radars \cite{gurbuz2021american,sang2018micro,kim2016hand,smith2021sterile}, Google ATAP's Project Soli is the only effort using mmWave radar as a musical interface, using gesture recognition and 1-D position estimation to control the parameters of audio synthesizers \cite{bernardo2017_o_soli_mio}. 

The novel framework presented in this article offers a major advancement for near-field mmWave hand-tracking and an accessible MATLAB software platform for further investigation into real-time mmWave HCI and algorithm innovation. 
2-D localization performance is considerably improved from past work \cite{joshi2015wideo} by employing a novel deep learning-based technique to improve the resolution beyond the theoretical limitations.

Existing work on contactless gesture control, such as gesture radar, commonly applies machine learning and deep learning techniques.
Gurbuz \textit{et al.} employ a multi-frequency radio-frequency (RF) sensor to recognize American sign language patterns with high accuracy using several machine learning techniques such as support vector machines (SVM), random forest, linear discriminant analysis, and k-nearest neighbors \cite{gurbuz2021american}.
Foot gestures are classified from EMG data using an SVM classifier \cite{maragliulo2019foot}. 
In \cite{sang2018micro}, Sang \textit{et al.} compare the classification rates from several techniques from traditional machine learning approaches such as hidden Markov models (HMMs) to state-of-the-art deep learning models including convolutional neural networks (CNNs), recurrent neural networks (RNNs), and end-to-end networks. 
On mmWave radar, \cite{kim2016hand} proposes a CNN classifier for dynamic gestures using micro-Doppler signatures, and \cite{smith2021sterile} employs a sterile technique to improve the classification rate of static, non-moving hand poses.
These previous efforts offer a sufficient solution to many classification applications where data are collected and used to determine the gesture or hand pose performed by a subject from a set of predefined classes.

In this article, however, we consider a distinctly separate issue and offer a deep learning-based signal processing solution.
Rather than classifying a sample from a set of classes, our proposed framework seeks to extract continuous spatial and temporal features from the hand's position and motion by improving the spatiotemporal image resolution.
To achieve this goal, we apply a novel fully convolutional neural network (FCNN) to preserve the geometry of the image and perform super-resolution for improved localization.
Radar signal processing using FCNNs is advantageous over other CNN techniques as it allows for data-driven "enhancement" rather than dimensionality reduction, as in classification.
Hence, rather than suffering from information loss, the regressive FCNN provides additional "context" learned during the training phase to enhance the radar data.
The enhanced data offer several advantages such as improved signal-to-noise-ratio (SNR), clutter removal, near-field image correction, aliasing suppression, and higher-resolution peaks.
In this article, traditional radar signal processing algorithms are shown to achieve considerable performance gains when applied to enhanced data.
In \cite{gao2018enhanced} and \cite{kim2020aziumth}, an FCNN and U-Net are employed to enhance the resolution of a radar image under far-field, plane-wave assumptions.
Our novel approach unifies FCNN-based super-resolution with near-field imaging, which requires more difficult spherical-wave compensation, on a small (8-channel) array and is shown to improve hand-tracking performance significantly.
To our knowledge, this article is the first documented effort towards near-field radar image super-resolution using an FCNN approach for improved localization.
Incorporating our enhancement FCNN in the signal processing chain enables fine motion tracking unattainable by existing techniques.
Additionally, a particle filter tracking algorithm is presented to further improve tracking robustness by employing the Doppler effect. 
Compared to prior work on gesture tracking using optical solutions \cite{sun2019visual,polfreman2011multi,jensenius2013kinectofon,han2014lessons,hantrakul2014implementations,nieto2013hand}, our approach offers fine hand-tracking using a single mmWave sensor offering higher depth resolution with superior privacy. 
This article proposes a novel hand-tracking method for musical interface by fusing spatiotemporal algorithms, deep learning-enhanced feature extraction, and robust position tracking algorithms. 
To aid further development and prototyping for real-time mmWave gesture applications, the entire software implementation is available by request to the corresponding author. 
To our knowledge, this proposed framework is the first openly available software package supporting real-time data streaming from a mmWave radar into MATLAB for streamlined signal processing and deep learning algorithm development. 

The remainder of this article is organized as follows. 
Section \ref{sec:radar_theory_for_musicians} provides an overview of the frequency modulated continuous wave (FMCW) radar signal model and feature extraction methods. 
In Section \ref{sec:the_radar_musical_instrument}, two robust tracking algorithms and estimation techniques are presented.
The system implementation is discussed in Section \ref{sec:system_design_and_implementation}, and results are shown in Section \ref{sec:results}. 
Section \ref{sec:discussion} provides a discussion of the performance, design constraints, and distinct advantages of the two tracking methods in Sections \ref{subsec:classical_gesture_tracking} and \ref{subsec:enhanced_gesture_tracking}, followed finally by conclusions.

\textit{Notation:} Throughout this article, vectors and matrices are set in boldface, using lowercase letters for vectors and uppercase letters for matrices. 
The superscripts $^T$ and $^*$ denote the transpose and conjugation operations, respectively. 
The all-ones and all-zeros vectors, of size $N \times 1$, are expressed as $\mathbf{1}_N$ and $\mathbf{0}_N$, respectively. 
The variable $c$ represents the free-space speed of light.
Finally, the multivariate Gaussian distribution with mean vector $\mathbf{\mu}$ and covariance matrix $\mathbf{\Sigma}$ is denoted as $G(\mathbf{\mu},\mathbf{\Sigma})$.
Spatial coordinates are treated as a continuous domain to support continuously distributed target scenes and time variables are modeled in discrete-time.

\section{Preliminaries of MIMO-FMCW Radar Signaling}
\label{sec:radar_theory_for_musicians}
\begin{figure}[h]
	\centering
	\includegraphics[width=0.35\textwidth]{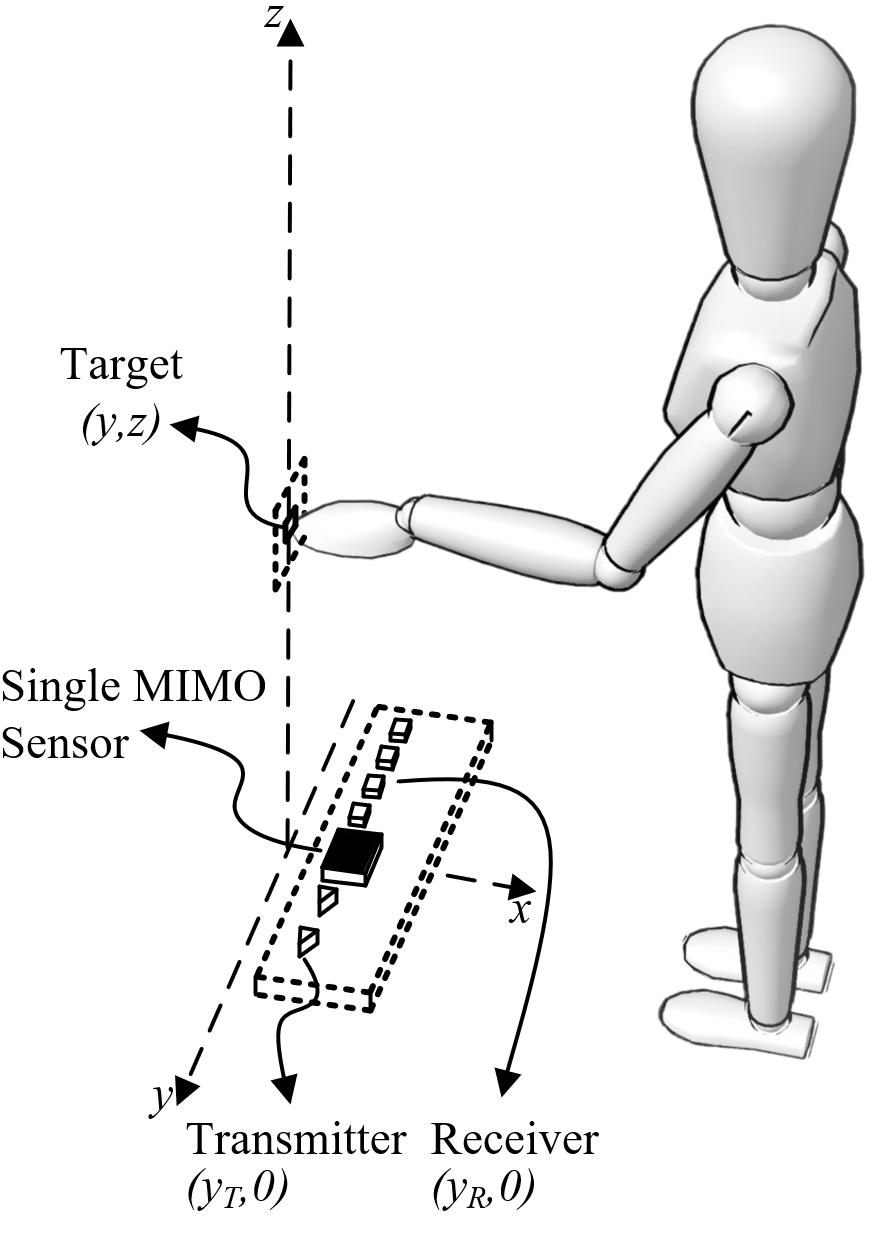}
	\caption{The imaging geometry, where the linear MIMO array faces vertically and the musician moves their hand throughout the $y$-$z$ plane.}
	\label{fig:MIMO_radar_musical_instrument_setup}
\end{figure}

In this section, we overview the propagation model for the FMCW radar chirp signal and examine the spatiotemporal features of a target in motion. 
The imaging scenario, as shown in Fig. \ref{fig:MIMO_radar_musical_instrument_setup}, consists of a multistatic linear MIMO array facing vertically. 
Orthogonality is leveraged in time by employing time-division-multiplexing MIMO (TDM-MIMO), wherein the transmitters are activated at separate time instances.
Throughout this article, the musician's hand is modeled as a point reflector located at the point $(y,z)$.

\subsection{MIMO-FMCW Signal Model}
\label{subsec:signal_model}
The FMCW chirp signal model is well documented in the literature \cite{TI:rao2017intro,smith2020nearfieldisar,yanik2019sparse} and is discussed in this section for reference and continuity throughout this article.
First, consider a single transmitter/receiver pair located at $(y_T,0)$ and $(y_R,0)$ in the $y$-$z$ plane, respectively, and an ideal point target with reflectivity $p$ located at $(y,z)$.
Assuming ideal propagation on a noiseless channel, the continuous-time signal can be modeled as
\begin{equation}
\label{eq:mimo_beat_signal_continuous}
    s(y_T,y_R,k) = \frac{p}{R_T R_R} e^{jk(R_T + R_R)},
\end{equation}
where $R_T$, $R_R$ are the distances from the transmitter and receiver to the point target, respectively, and $k = 2\pi f/c$ is the instantaneous wavenumber.
The frequency of the chirp $f = f_0 + Kt$ increases linearly against time with slope $K$ and starting frequency $f_0$.
The continuous-time signal (\ref{eq:mimo_beat_signal_continuous}) is sampled with sampling frequency $f_S$ by the radar analog-to-digital converter (ADC) and can be written in discrete time as
\begin{equation}
	\label{eq:mimo_beat_signal}
	s(y_T,y_R,n_k) = \frac{p}{R_T R_R} e^{j(k_0 + \Delta n_k)(R_T + R_R)},
\end{equation}
where $n_k$ is the wavenumber index, $k_0 = 2\pi f_0/c$ is the starting wavenumber corresponding to the starting frequency $f_0$, and $\Delta = 2\pi K /(c f_S)$ is the wavenumber step size.

To ease the subsequent signal processing, it is desirable to approximate the multistatic MIMO beat signal, represented in (\ref{eq:mimo_beat_signal}) as its corresponding monostatic equivalent using the approximation developed in \cite{yanik2020development} as
\begin{equation}
	\label{eq:mult-to-mono}
	\hat{s}(y',n_k) = s(y_T,y_R,n_k) e^{-j(k_0 + \Delta n_k)\frac{d_y^2}{4z_0}},
\end{equation}
valid only for small $d_y$, the distance between the transmitter and receiver elements, where $z_0$ is a reference plane typically given as the center of the target scene. 
Taking $y'$ as the locations of the virtual elements located at the midpoints between each transceiver pair and $R$ as the corresponding distance from each virtual element to the point reflector, the resulting monostatic beat signal approximates to
\begin{equation}
	\label{eq:mono_beat_signal}
	\hat{s}(y',n_k) \approx \frac{p}{R^2}e^{j2(k_0 + \Delta n_k)R}.
\end{equation}
From (\ref{eq:mono_beat_signal}), the spatial location, $(y,z)$, of the target is embedded in the radar beat signal, in the form of the radial distance $R$.

\subsection{Doppler Radar Signal Processing}
\label{subsec:fmcw_doppler_radar}
The relative velocity of a target can be extracted from the beat signal expressed in (\ref{eq:mono_beat_signal}) by exploiting the Doppler effect.
As discussed in \cite{winkler2007range}, by transmitting a series of chirp waveforms at a known pulse repetition interval (PRI), $T_{PRI}$, the velocity of a moving target can be identified as the frequency component along the chirp index dimension given by
\begin{equation}
	\label{eq:doppler_final}
	\hat{s}(y',n_k,n_c) = \frac{p}{R^2} e^{j(2(k_0 + \Delta n_k)R + \frac{4\pi v T_{PRI}}{\lambda_0}n_c)},
\end{equation}
where $R$ is the initial range of the target, $v$ is the velocity of the target, $\lambda_0$ is the wavelength corresponding to $f_0$, and $n_c$ is the chirp index,

Thus, the beat signal sampled across time is a 2-D complex sinusoidal with frequencies corresponding to the range and velocity of the target on the first and second dimensions, respectively. Subsequently, to extract the range and velocity, traditional methods perform a 2-D fast Fourier transform (FFT) over a matrix whose rows or columns consist of subsequent chirps. 

\subsection{Range Migration Algorithm Image Reconstruction}
\label{subsec:rma}
To achieve high-fidelity 2-D localization, we employ the range migration algorithm (RMA) over traditional range-angle FFT methods \cite{TI:rao2017intro, kim2020aziumth}, whose localization accuracy is known to be inferior \cite{kim2018joint}. 
The primary goal of the RMA is to reconstruct the target scene's reflectivity function, $p(y,z)$.
For a distributed target, the beat signal can be modeled as the superposition of the backscattered signal at every point in the scene, neglecting the amplitude terms, as
\begin{equation}
	\label{eq:rma_1}
	\hat{s}(y',n_k) = \iint p(y,z) e^{j2(k_0 + \Delta n_k)R}dydz.
\end{equation}

This target model assumes a spatially distributed target whose reflectivity only depends on spatial location and neglects any frequency dependence of the reflectivity function. 
Inverting (\ref{eq:rma_1}) using the method of stationary phase, the reflectivity function, $p(y,z)$, can be estimated efficiently by
\begin{equation}
	\label{eq:rma_summary}
	\hat{p}(y,z) = \text{IFT}_{\text{2D}}^{(k_y,k_z)}\left[ \mathcal{S}\left[ \text{IFT}_{\text{1D}}^{(y')}[\hat{s}^*(y',n_k)] \right] \right],
\end{equation}
where $\mathcal{S}[\bullet]$ is the Stolt interpolation operation \cite{yanik2020development} and $\text{FT}[\bullet]$, $\text{IFT}[\bullet]$ are the forward and inverse Fourier transform operators, respectively. 
To avoid aliasing in the image sampling criteria must be considered \cite{yanik2019sparse}. Spatial resolution along the $y$ and $z$-directions are constrained by the physical and device limitations and are expressed as 
\begin{equation}
    \label{eq:y_res}
    \delta_y = \frac{\lambda_c z_0}{2 D_y},
\end{equation}
\begin{equation}
    \label{eq:z_res}
    \delta_z = \frac{c}{2 B},
\end{equation}
where $\lambda_c$ is the wavelength corresponding to the frequency at the center of the chirp sweep, $D_y$ is the aperture size along the $y$-direction, and $z_0$ is the center of the imaging scene \cite{yanik2019sparse}.

After the 2-D reflectivity function of the target scene is recovered, the hand position is estimated subsequently as
\begin{equation}
\label{eq:location_estimate}
    \{\hat{y},\hat{z}\} = \arg \max_{\{y,z\}} \hat{p}(y,z).
\end{equation}

Further, the aforementioned Doppler principle can be leveraged to extract the velocity of the target by Fourier analysis over successive chirps. 
To optimally exploit the deep learning framework discussed in Section \ref{subsubsec:improved_2d_position_esitmation_by_FCNN} and reduce the required computation complexity, the velocity is extracted after the RMA is performed and hand location is estimated.

As evident in (\ref{eq:doppler_final}), the velocity is decoupled from the wavenumber index and is the scaled frequency component along the chirp index dimension. 
As a result, the phase term corresponding to the velocity is preserved in the reconstructed image, $\hat{p}(y,z)$. 
Therefore, the velocity profile can be obtained by performing an FFT across the chirp index, $n_c$, dimension of the recent images.
Rather than performing the FFT across the 3-D array, $\hat{p}(y,z,n_c)$, we perform the FFT over the slice of the image corresponding to the estimated position, $\hat{y}$, yielding the velocity profile along the $z$-direction, where $n_d$ is the velocity index, as

\begin{equation}
	\label{eq:doppler_fft}
	\hat{d}(z,n_d) = \text{FFT}_{\text{1D}}^{(n_c)} \left[ \hat{p}(y,z,n_c) \biggr\rvert_{y = \hat{y}} \right].
\end{equation}

Finally, the velocity can be estimated from (\ref{eq:doppler_fft}) using video pulse integration by 
\begin{equation}
	\label{eq:velocity_doppler}
	\hat{v}_d = \arg \max \sqrt{\int |\tilde{d}(z,n_d)|^2 dz}.
\end{equation}
The velocity computed by this method is referred to as the Doppler velocity.
The recovered velocity using this approach is limited by the timing and physical constraints between $[-\frac{\lambda_0}{4T_{PRI}},\frac{\lambda_0}{4T_{PRI}}]$.
Later, the Doppler velocity is employed to improve the tracking performance using the Doppler corroborated particle filter.

\section{Spatiotemporal Imaging on mmWave Radar}
\label{sec:the_radar_musical_instrument}
In this section, we present the methods for our proposed imager capable of high accuracy hand-tracking for HCI. 
The contribution of this article is the advancement in algorithm performance for 2-D localization by utilizing both the novel super-resolution FCNN and the proposed tracking algorithm. 
While we will investigate the application of such algorithms as an NMI, our mmWave radar-based sensing algorithms can be applied to a host of HCI problems.

It is important to note that this work is not intended to compete with the computational efficiency of embedded HCI solutions and existing musical interfaces. 
Rather, the main contributions of this article are novel algorithms for super-resolution spatiotemporal hand-tracking and a freely-downloadable platform to increase accessibility and encourage further research in this arena. 
As such, we will focus primarily on the development of the algorithms and their localization performance.
Discussions on performance and implementation issues are considered secondary and are addressed in Sections \ref{sec:system_design_and_implementation} and \ref{sec:discussion}.

\subsection{Classical Spatiotemporal Feature Extraction Techniques}
\label{subsec:classical_gesture_tracking}
In this section, we introduce the simple approach to spatiotemporal sensing for contactless musical instrument interface.
While our system generally tracks the 2-D position and velocity of the user's hand, we have identified three underlying features to achieve fine control of the musical interface: range, cross-range oscillation, and velocity. 
By the geometry given in Fig. \ref{fig:MIMO_radar_musical_instrument_setup}, we define the range as the position of the hand along the $z$-axis, i.e. the vertical displacement between the sensor and the user's hand. 
Similarly, cross-range is defined as the position of the hand along the $y$-axis.
Subsequently, cross-range oscillation is the rate at which the hand oscillates in the cross-range direction.
Velocity is given by the velocity of the hand with respect to the range $z$-axis. 
These parameters are selected such that the output musical interface is controlled primarily by the range of the musician's hand and secondarily by the cross-range oscillation and velocity.
However, these parameters can be assigned by the user based on preference using the MIDI interface, as discussed in Section \ref{sec:system_design_and_implementation}.
Throughout the remainder of this article, we will refer to these parameters as features extracted from the radar beat signal.

Under the simple gesture tracking regime, the 2-D location and velocity $(\hat{y},\hat{z},\hat{v}_d)$ are extracted from the reconstructed image and buffer of recent images using (\ref{eq:location_estimate}) and (\ref{eq:velocity_doppler}). 
In the next section, the three parameters extracted from the raw data are treated as a vector called the noisy measurement vector $\mathbf{r}$. 
In the optimal scenario, the bandwidth, antenna array size, and SNR are quite large, tending towards infinity.
For the case of an $8$ channel automotive mmWave radar and a human hand, the bandwidth is limited ($4$ GHz), the antenna array size is small ($D_y = 2\lambda_c$), and the reflectivity of the hand is not high compared to the noise level. 
As a result, simply extracting the maximum from the reconstructed RMA images yields sporadic location and velocity estimates. 
Even in the ideal case, the spatial resolution of our system along the $y$ and $z$-directions is $\delta_y = 7.5$ cm and $\delta_z = 3.75$ cm, respectively.
Several other factors are not taken into account in the classical, direct tracking method including beam-pattern, residual phase errors, and antenna coupling. 
All these limitations and non-idealities in the imaging scenario degrade the image and result in noisy location and velocity estimates; however, many of these issues analytical forms and cannot be solved directly by classical methods. 
To address these issues, we present a novel data-driven approach employing an FCNN for super-resolution and image enhancement. 

\subsection{FCNN-Based Super-Resolution Feature Extraction and Particle Filter Tracking Methods}
\label{subsec:enhanced_gesture_tracking}
In this section, we improve upon the simple tracking techniques to overcome noise and foundational non-idealities in the imaging scenario, yielding a much-improved user experience. 
The concepts demonstrated in this section are applicable for many tracking and high-resolution imaging applications beyond the scope of musical interfaces.

To improve the tracking robustness of the proposed musical interface, we adopt the well-known particle filter \cite{garcia2013tracking} and present a novel modification. 
While traditional methods such as the extended Kalman filter (EKF) employ a motion model, our implementation of the particle filter bypasses the need for a deterministic motion model.
The particle filter is selected for this application as other traditional approaches have demonstrated poor tracking performance in our experimentation, yielding either sporadic localization or overly damped, sluggish estimation.
Additionally, the particle filter is advantageous as it can track non-linear dynamics and does not require prior knowledge of the motion model or noise parameters for robust localization.
Our proposed extension of the particle filter introduces a novel particle resampling and weight calculation procedure.
These modifications do not significantly alter the existing particle filter framework but are included to demonstrate their viability in hand-tracking with radar and adequately document our complete software implementation.
Furthermore, the stability of the particle filter algorithm has been investigated elsewhere and will not be addressed in this article \cite{le2004stability}.

In our modification of the particle filter, the control input is a weighted movement towards the newest measurement. 
To demonstrate our proposed algorithm, consider the case of simultaneous location estimation along the $y$ and $z$-directions.
The new noisy measurement vector, $\mathbf{r}$, has two elements, the newest estimates of the location, $\hat{y}$ and $\hat{z}$, which are extracted by the methods described in the prior section.
For 2-D localization, $\mathbf{X}_n$ is a matrix of size $N \times 2$, whose rows are the $(y,z)$ coordinates of each particle at time index $n$, where $N$ is the number of particles, and $\mathbf{w}_n$ is the vector of weights corresponding to each particle.
The estimates of the 2-D location (also known as the estimated states) form the vector $\mathbf{s}_n$.

Before executing the iterative algorithm, the initial particle states matrix, $\mathbf{X}_{0}$, and initial weights vector, $\mathbf{w}_{0}$, are initialized with random locations throughout the region of interest (ROI) and uniform weights, respectively. 
At each iteration, the particle filter receives $\mathbf{r} = [\hat{y},\hat{z}]^T$, the newest location estimates; $\mathbf{\hat{X}}_{n-1}$, the previous particle filter locations, $\mathbf{X}_{n-1}$, sampled using weights $\mathbf{w}_{n-1}$; $\mathbf{a} = [a_y, a_z]^T$, the weighting vector whose two elements provide weight to the noisy estimates $\hat{y}$ and $\hat{z}$, respectively; $\mathbf{s}_{n-1}$, the vector of previously estimated states, $\mathbf{\Psi} = [\bm{\psi}_1, \bm{\psi}_2]$, a matrix consisting of two random vectors from the distribution $G(\mathbf{0}_N,\mathbf{\Sigma_\psi})$; and $\mathbf{\Sigma}_w$, the covariance matrix for the weight distribution.

\begin{algorithm}[h] 
\renewcommand{\thealgocf}{}
	\label{algo:particle_filter}
	\caption{Modified Particle Filter Algorithm}
	\SetAlgoLined
	\SetKwInOut{Input}{input}
	\SetKwInOut{Output}{output}
	
	\Input{$\mathbf{r} = [\hat{y},\hat{z}]^T, \mathbf{\hat{X}}_{n-1}, \mathbf{a}, \mathbf{s}_{n-1}, \mathbf{\Psi}, \mathbf{\Sigma}_w$}
	\Output{$\mathbf{s}_n = [\tilde{y},\tilde{z}]^T$}
	
	$\mathbf{X}_n \xleftarrow[]{} \mathbf{\hat{X}}_{n-1} + \mathbf{1}_N \mathbf{a}^T (\mathbf{r} - \mathbf{s}_{n-1}) + \mathbf{\Psi}$ 
	
	$\mathbf{w}_n \xleftarrow[]{} e^{-\frac{1}{2}(\mathbf{X}_n-\mathbf{s}_{n-1})^T\mathbf{\Sigma}_w^{-1}(\mathbf{X}_n-\mathbf{s}_{n-1})}$
	
	$\mathbf{s}_n \xleftarrow[]{} \frac{1}{\mathbf{1}_N^T \mathbf{w}_n} \mathbf{X}_n^T \mathbf{w}_n$ 
\end{algorithm}

Proper handling of the key steps, (step 1) resampling of the particle states and (step 2) computing new weights, is essential to effectively implement our proposed particle filter algorithm. 

\begin{figure}[h]
	\centering
	\includegraphics[width=0.65\textwidth]{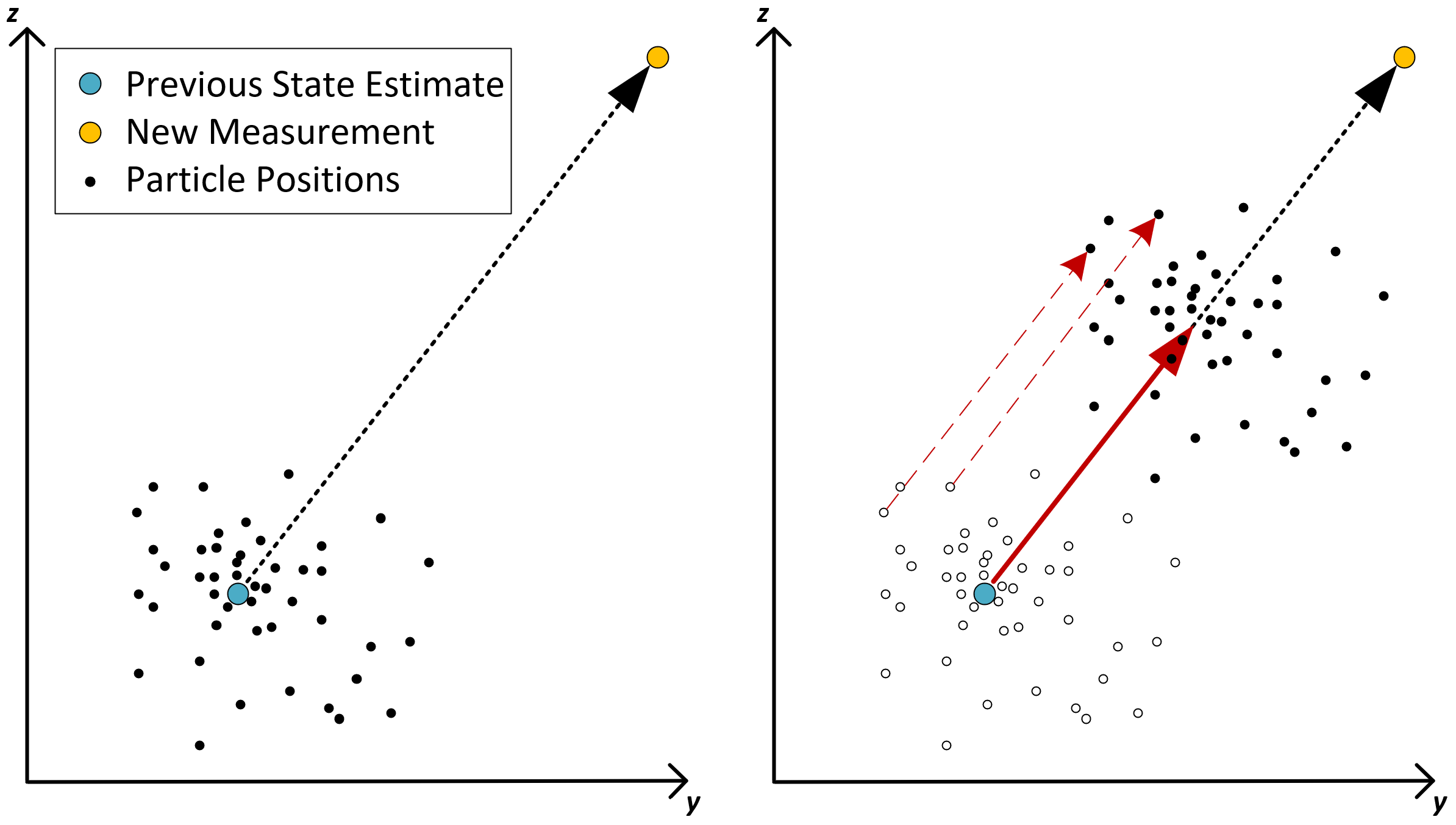}
	\caption{A visual example of the modified particle filter algorithm resampling process. The particle locations are resampled by a shift transformation towards the new measurement according to the weight vector $\mathbf{a}$, where $a_y = a_z = 0.5$.}
	\label{fig:particle_filter}
\end{figure}

The particle resampling process involves moving the particles towards the new measurement by a specified weight.
The size of $\mathbf{a}$, $\mathbf{r}$, and $\mathbf{s}_n$ can be varied depending on the number of parameters to be tracked by the particle filter.
Hence, the new measurements do not dominate the motion tracking but have a weighted influence on the localization procedure. 
Fig. \ref{fig:particle_filter} demonstrates the resampling process with $a_y = a_z = 0.5$.
Note that before computing the new weights, particle diffusion is performed by adding the perturbation term $\mathbf{\Psi}$.

The new weights are computed from a multivariate Gaussian distribution with the previously estimated states, $\mathbf{s}_{n-1}$, as the mean vector and a predefined covariance matrix $\mathbf{\Sigma}_w$.
Therefore, particles closer to the previously estimated state are assigned a higher weight than those farther away. 
This results in a tendency towards small changes in the state estimations while monitoring for movement from the current position. 
To our knowledge, the resampling and weight calculation steps proposed in this section offer a novel implementation of the particle filter. 
For many applications requiring precise and consistent localization and motion tracking on mmWave radar, our modified particle filter algorithm is an ideal fit as it tends to a steady-state estimation of the states but remains active in monitoring the noisy sensor input.

\subsubsection{Doppler-Corroborated Real-Time Weighting}
In this section, we present a dynamic weighting technique for updating $\mathbf{a}$ in real-time by exploiting the dependence between position and velocity.
Our approach considers corroboration between the Doppler velocity estimate and the velocity estimated from the range samples as a measure of the new measurement's reliability. 
Thus, the dependability of the Doppler velocity can improve tracking of the target position along the range ($z$) dimension even in the presence of noisy position estimates. 
After the Doppler velocity is calculated by (\ref{eq:velocity_doppler}), the recent range estimates are used to calculate the sample velocity ($\hat{v}_s$) by the least-squares estimator as
\begin{gather}
	\hat{v}_s = \frac{N_z T_{PRI} \sum_m (\mathbf{z}^{(m)} m) - T_{PRI} \sum_m \mathbf{z}^{(m)} \sum_m m}{N_z \sum_m (\mathbf{z}^{(m)})^2 - \left( \sum_m \mathbf{z}^{(m)} \right)^2},
\end{gather}
where $\mathbf{z}^{(m)}$ is the $m^{\text{th}}$ element of the vector of recent $\hat{z}$ estimates, $\mathbf{z}$, with $\mathbf{z}^{(N_z-1)}$ being the most recent. 

The difference between the Doppler estimated velocity and sample estimated velocity is computed as $\Delta_v = |\hat{v}_d - \hat{v}_s|$ and used in the reward function (\ref{eq:velocity_cost_function}) to update the weight placed on the new noisy measurement in real-time. 

\begin{equation}
	\label{eq:velocity_cost_function}
	a_z (\Delta_v)=\begin{cases}
		a_{z,0} \cos \left(\frac{2\pi T_{PRI} \Delta_v}{\lambda_0}\right) \quad &\text{if} \, \Delta_v \leq \frac{\lambda_0}{4T_{PRI}} \\
		0 \quad &\text{if} \, \Delta_v > \frac{\lambda_0}{4T_{PRI}} \\
	\end{cases}
\end{equation}

When the sample velocity is close to Doppler velocity, i.e. $\Delta_v$ is small, the reward function is close to $a_{z,0}$. 
Hence, the new measurement is corroborated by the reliable Doppler velocity and weighted accordingly. 
Outliers and erroneous measurements contradicting the Doppler velocity are given less importance during the particle resampling process. 
To implement the Doppler corroborated particle filter, $\mathbf{a} = [a_y, a_z(\Delta_v)]^T$  is dynamically updated by (\ref{eq:velocity_cost_function}) at each iteration of our proposed particle filter algorithm.

\subsubsection{Improved 2-D Position Estimation by Enhancing FCNN}
\label{subsubsec:improved_2d_position_esitmation_by_FCNN}
The modified particle filter algorithm improves the tracking consistency and smoothness; however, several issues such as instrumentation delay, ambient/device noise, multistatic effects, and non-spherical beam patterns remain unaddressed and degrade tracking performance. 
To overcome these non-idealities, we present a novel FCNN-based technique for image enhancement that improves the 2-D position estimation, subsequent tracking accuracy, and Doppler spectrum SNR. 
Compared to prior FCNN synthetic aperture radar (SAR) techniques employing far-field assumptions and trained on synthetically generated data \cite{gao2018enhanced}, our enhancement FCNN method operates on near-field images, improves localization even with a small aperture, and is trained using a novel technique allowing the network to learn the environment and device noise, near-field beam pattern, and multistatic effects.

To train the enhancement FCNN, we construct a dataset consisting of both real human hand data and synthetically generated data.
Real hand data are collected by capturing frames while the user holds their hand at known locations relative to the device and synthetic data are used to supplement the training set. 
Each synthetic sample is generated by simulating a MIMO beat signal using (\ref{eq:rma_1}) with one ideal point target located at a known location and additive real device noise, collected from the radar. 
The simulated locations are randomized to uniformly cover the ROI.
Both the real and synthetic data are used as features in the FCNN training process, thus enabling the network to fit the non-ideal beam pattern, real multipath and multistatic effects, the empirical reflection of a human hand, device and ambient noise, and hand positions throughout the ROI. 

To train the image-to-image regression FCNN, each training feature (real or synthetic image) must correspond to a ground truth label.
The ground truth label images are synthetically generated by the model
\begin{equation}
	\label{eq:fcnn_expected_image}
	\mathcal{I}(y,z) = e^{-(y - y_0)^2/\sigma_y^2 -(z - z_0)^2/\sigma_z^2}
\end{equation}
where the width of the expected target located at $(y_0,z_0)$ is dictated by $\sigma_y$ and $\sigma_z$ in the $y$ and $z$ dimensions respectively, yielding resolutions of $1.18\sigma_y$ and $1.18\sigma_z$ according to the $3$ dB beamwidth definition \cite{gao2018enhanced}. 
Each label is generated using the requisite knowledge of the location of the human hand or target of each feature image. 
During training, the FCNN learns the highly nonlinear relationship between distorted, blurred RMA images and the ideal images generated using (\ref{eq:fcnn_expected_image}).
Our novel training technique results in a robust and generalizable FCNN that improves image SNR and localization by fitting to the non-ideal imaging constraints. 
Further, the trained network enables localization precision beyond the physical limitations of the device improving tracking performance significantly.
FCNN training is discussed in Section \ref{subsec:enhanced_tracking_implementation} and results are presented and discussed in Section \ref{subsec:enhanced_gesture_tracking_results}.

Additionally, by isolating the peak corresponding to the human hand, clutter and phase noise at other positions are mitigated thereby improving the Doppler spectrum SNR and subsequent velocity estimation.
Thus, the FCNN enhances both the spatial and temporal features extracted from the radar beat signal before the particle filter.
Uniting the proposed particle filter and enhancement FCNN, the range, cross-range oscillation, and velocity are robustly tracked by our novel algorithms and mapped to musical interface controls.

\section{System Design and Implementation}
\label{sec:system_design_and_implementation}
In this section, we present the system implementation for both the classical tracking techniques and our novel super-resolution feature extraction and tracking algorithms discussed in the previous section.

\subsection{Hardware and Software Implementation}
\label{subsec:hardware_setup_and_challenges}
The hardware employed in the proposed system consists of a Texas Instruments (TI) AWR1243 automotive radar in conjunction with a DCA1000EVM real-time data capture adapter. 
The TI radar is a MIMO-FMCW mmWave radar with an operating bandwidth of $4$ GHz and a center frequency of $79$ GHz. 
In this research, we utilize the linear MIMO array consisting of $2$ transmit antenna (TX) elements, separated by $2\lambda_c$, and $4$ receive antenna (RX) elements, separated by $\lambda_c/2$.
The resulting virtual array has 8 equally spaced virtual elements separated by $\lambda_c/4$ \cite{TI:rao2017intro}. 
The calibration methodology discussed in \cite{yanik2020development} is adopted to mitigate range bias, constant phase errors, and instrumentation delay. 
In this process, data are captured from a corner reflector at a known location and used to identify range bias and phase offsets among the antennas.
Unlike an optical or infrared calibration, this process is invariant of lighting and temperature constraints as well as user hand sizes, etc.
Thus, the one-time calibration applies to a variety of environments and users.

The software platform for signal processing, visualization, and machine learning is written in MATLAB. 
Despite its inferior computational efficiency compared to other languages, MATLAB is employed to provide an accessible platform for researchers to engage with this work and rapidly prototype custom real-time algorithms using our custom tools.
Once the algorithms are validated on a PC, they can be implemented onto such embedded devices for optimized application-specific usage.
For a positive user experience as a musical interface, latency and timing issues must be taken into account, and are discussed in Section \ref{sec:discussion}.

\begin{figure}[ht]
	\centering
	\includegraphics[width=0.8\textwidth]{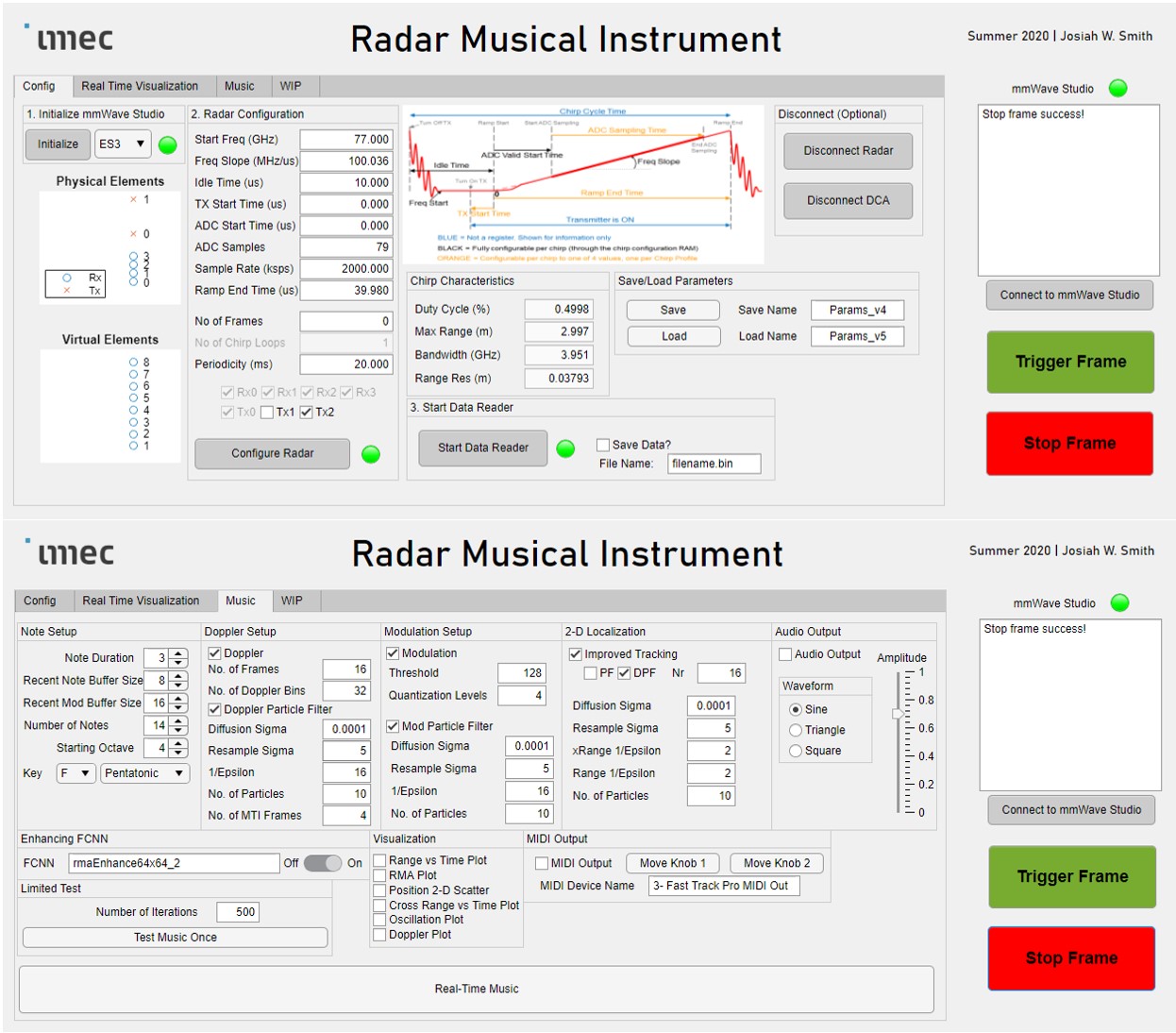}
	\caption{Interactive MATLAB GUI: single interface for the user to setup the radar device, specify parameters for the tracking and music generation, and start and stop device captures.}
	\label{fig:matlab_gui}
\end{figure}

\subsection{Real-Time Data Retrieval and Interactive MATLAB User Interface}
\label{subsec:real_time_data_retrieval_and_MATLAB_interface}
To stream the data from the device into MATLAB, a custom UDP interface software is written. 
This routine is implemented efficiently in C++ and is capable of receiving the sequential UDP packets, organizing the packets to form each chirp, and providing the data to MATLAB over shared memory. 

A custom interactive MATLAB graphical user interface (GUI), shown in Fig. \ref{fig:matlab_gui}, is written to serve as the single user interface for our framework. 
The MATLAB GUI interfaces with TI mmWave Studio \cite{TI:mmWave_Studio} to control the hardware setup and initializes the UDP interface, bypassing the need for user setup outside our GUI.
The radar continuously captures and streams data into MATLAB using the fully integrated implementation. 
While MATLAB does not offer the computational speed necessary for real-time system implementation, it is capable of completing the data capture, signal processing, deep learning, visualization, and signal output at around $250$ Hz, from our experimentation. 
In the early prototyping phase, we consider this throughput sufficient for investigating the performance of the super-resolution tracking algorithms and a simple musical interface.

Using our proposed methods, the software extracts high-resolution spatiotemporal features of the user's hand and maps them to corresponding output using either a built-in audio output tool or the included MIDI output.
The custom MATLAB GUI provides an accessible option for investigating and demonstrating our methods as well as an open-source platform to stimulate further collaborative investigation by the multimedia and radar communities. 

As previously mentioned, the primary mechanism to control the output of the proposed musical interface is the range ($z$-position) of the user's hand.
Using the built-in audio output tool and the MIDI output, the range of the user's hand controls the note selection directly. 
Unlike the Theremin, which allows for continuous note selection, our interface quantizes the user input into predefined subregions corresponding to notes defined by the user.
To play the desired note, the user must move their hand vertically to the position corresponding to that note. 
The subregions and allowed notes can be programmed by the user in the interactive MATLAB GUI.
Continuous control, similar to the Theremin, can be easily implemented using our user interface.
However, true continuous pitch control, as with an analog Theremin, is achievable only to a certain extent as discretization is required at some point in a digital software.
We have selected discrete pitch control as the default for our framework for simple note selection to promote accessibility to numerous fields.
The Theremin is commonly regarded as one of the most difficult instruments to control given its continuous note selection and requires extensive experience to manage even simple musical phases with proper intonation.
Our solution allows advanced musicians to enable continuous pitch control while offering a simple interface for less-experienced users.
Thus, our platform is accessible to many users providing an enjoyable experience and enabling less-effortful demonstration without expert musicianship.
Similarly, the secondary parameters, cross-range oscillation, and velocity can be adjusted by the user by oscillating their hand back-and-forth in the $y$-direction or moving to the next note with a high or low velocity.
The built-in audio output tool employs the cross-range oscillation to control a vibrato effect (low-frequency modulation of the audio signal).
Thus, using this tool, the user can select the desired note by varying the range and perform vibrato at a desired rate by oscillating their hand at the same rate.
Alternatively, the MIDI output tool provides the cross-range oscillation and velocity as MIDI parameters to be specified by the user in a virtual instrument environment connected to the MIDI output of our musical interface.
Hence, our proposed algorithms are implemented to operate similarly to a MIDI keyboard with the hand range controlling the note selection and cross-range oscillation and velocity acting as MIDI parameters for the user to assign.

\begin{figure}[h]
	\centering
	\includegraphics[width=3.5in]{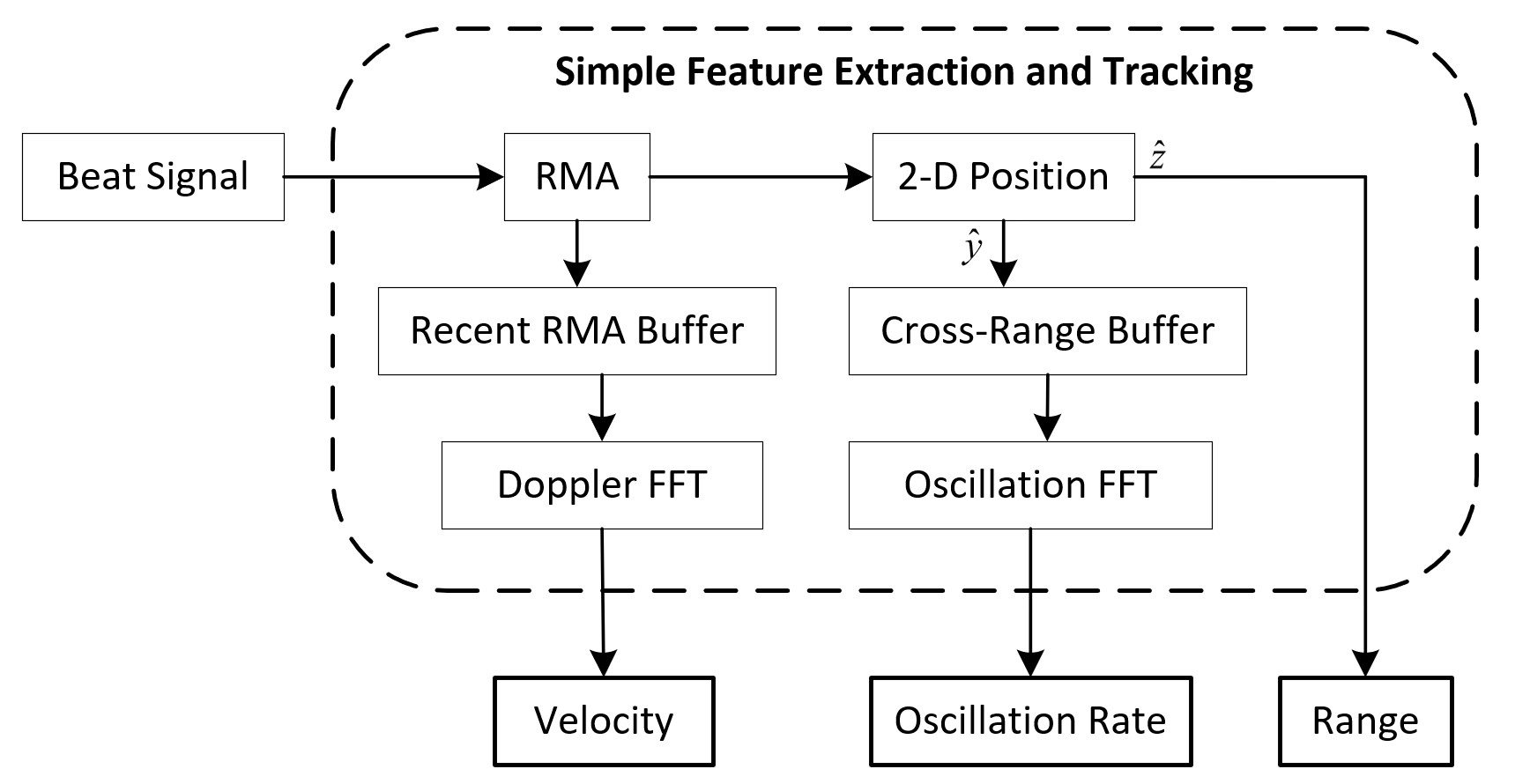}
	\caption{Simple tracking signal processing chain. After RMA is performed on the beat signal, features are extracted directly from the raw RMA image.}
	\label{fig:simple_signal_chain}
\end{figure}

\subsection{Simple Feature Extraction and Tracking Algorithm Signal Processing Chain}
\label{subsec:simple_tracking_implementation}
The signal processing chain for the simple feature extraction and tracking method is shown in Fig. \ref{fig:simple_signal_chain}. 
The beat signal is loaded into MATLAB where the preprocessing discussed in the previous section is performed (RMA and peak finding), and the user inputs (2-D location and velocity) are converted into audio or MIDI output by extracting the spatiotemporal features using (\ref{eq:location_estimate}) and (\ref{eq:velocity_doppler}).
In this article, the location and velocity of the user's hand are used for musical gestural interface; however, our novel algorithms can easily be applied to many different HCI applications and even for 3-D localization, provided a sufficient 2-D array.
The reconstructed RMA image and raw feature extracted by the classical techniques can be utilized by the particle filter algorithm and super-resolution FCNN to improve the tracking performance.

\subsection{Super-Resolution Framework - Training FCNN and Implementing Particle Filter Algorithm}
\label{subsec:enhanced_tracking_implementation}

\begin{figure}[ht]
	\centering
	\includegraphics[width=0.6\textwidth]{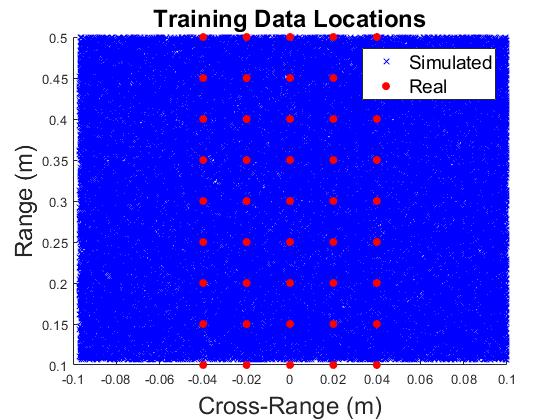}
	\caption{Locations of the training data used to train the enhancement FCNN. Real data (red) are collected by keeping the hand static at known locations. Simulated data (blue) are generated by choosing locations randomly from the continuous ROI.}
	\label{fig:fcnn_training_locations}
\end{figure}

To implement our super-resolution feature extraction and tracking framework, the super-resolution FCNN must be first trained.
The enhancement FCNN is trained using both real data from a human hand and simulated data corrupted by additive real radar noise. 
The FCNN is trained using $65536$ simulated and $23040$ real human hand RMA images as the input and output images with $\sigma_y = \sigma_z = 1$ mm resulting in cross-range and range resolutions of $1.18$ mm. 
Each simulated sample is generated at a random location in the ROI $y \in [-0.1,0.1]$, $z \in [0.1,0.5]$.
The synthetic data cover the entire ROI allowing the network to generalize well to location while learning the non-idealities of the imaging scheme.
$512$ samples of a real hand are collected at each of the $45$ locations throughout the ROI as shown in Fig. \ref{fig:fcnn_training_locations}.
For both the synthetic samples and real human samples, corresponding ground truth images are generated using (\ref{eq:fcnn_expected_image}) and used as training labels.
Thus, the training set is comprised of features consisting of real and simulated data and labels consisting of the ideal expected response at each known location.

\begin{figure}[h]
	\centering
	\includegraphics[width=0.75\textwidth]{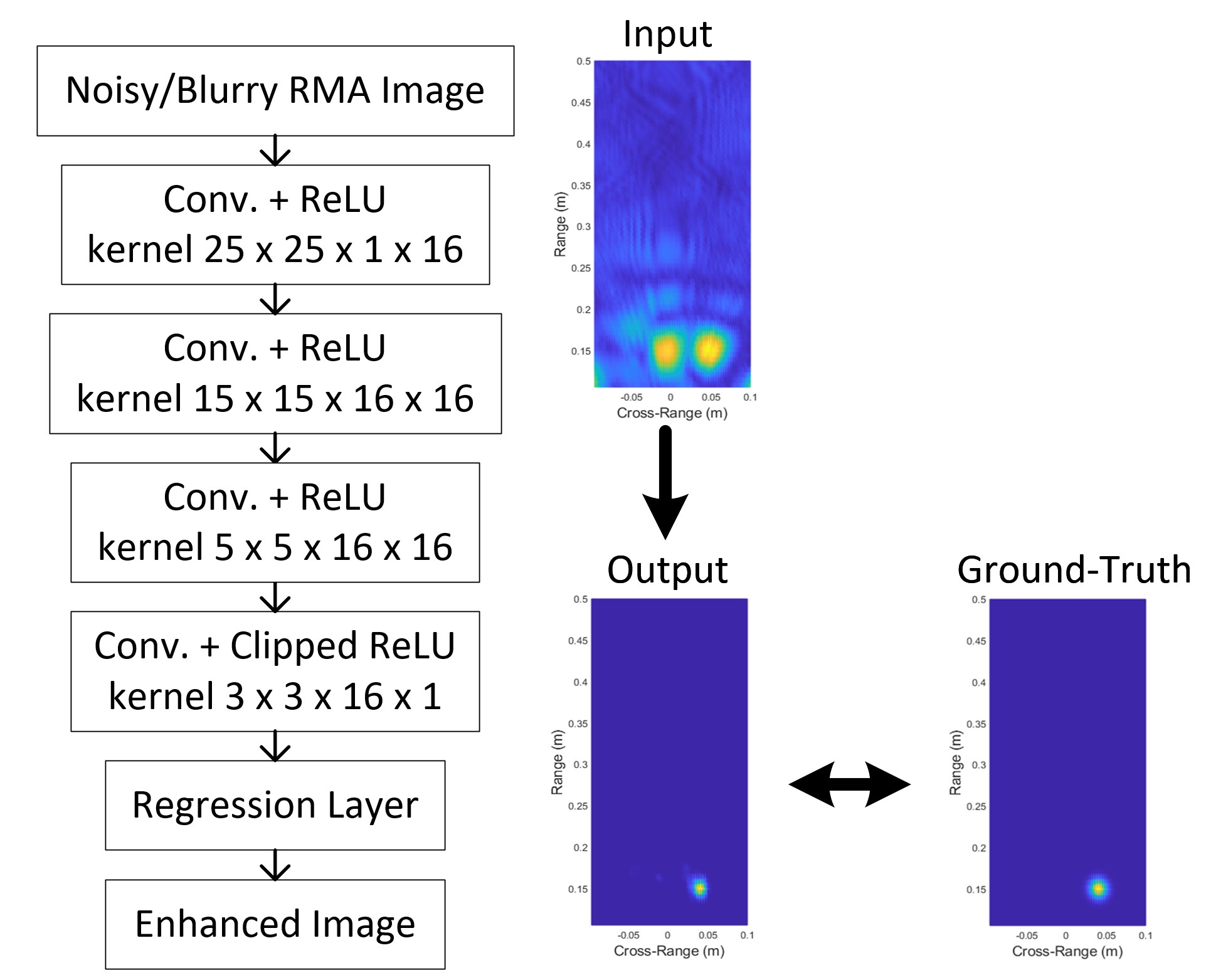}
	\caption{Network topology of the super-resolution enhancement FCNN. The selected kernel and layer sizes are capable of adequately learning the non-ideal shape of the distorted RMA image while maintaining high computational efficiency for real-time implementation.}
	\label{fig:fcnn_architecture}
\end{figure}

The architecture of the proposed enhancement FCNN is shown in Fig. \ref{fig:fcnn_architecture}.
The network consists of four convolution layers of decreasing kernel size each followed by a nonlinear Rectified Linear Unit (ReLU) layer. Each convolutional layer is zero-padded such that the output is identical in size to the input. 
Training the network for 100 epochs takes 5 hours on a machine with a single NVIDIA GTX1080TI graphics card. 
Other network architectures and training durations are investigated, but this combination yields high performance while offering real-time efficiency.

\begin{figure}[ht]
	\centering
	\includegraphics[width=0.75\textwidth]{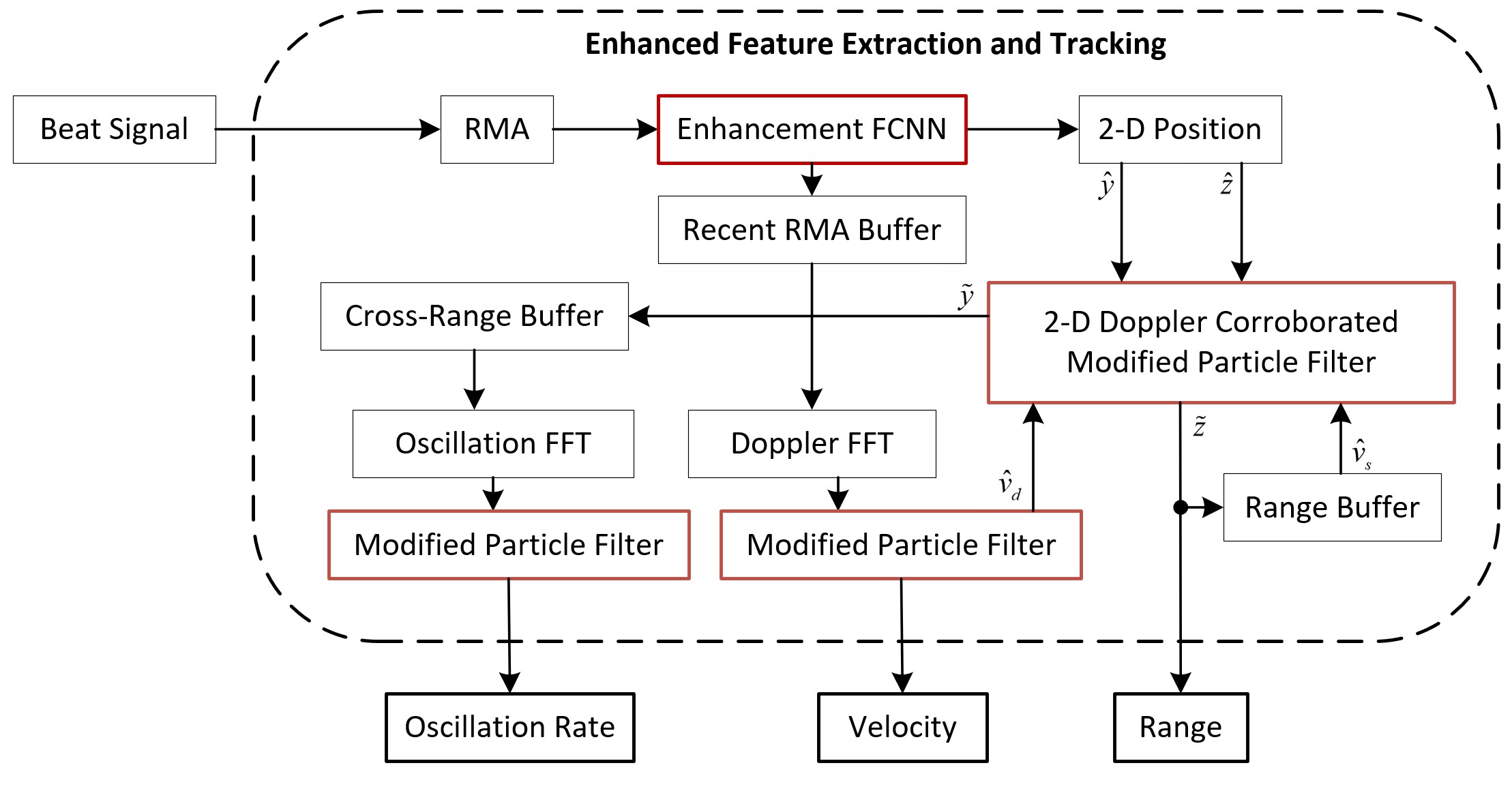}
	\caption{Enhanced tracking signal processing chain. Key elements to the enhanced methods are highlighted in red.}
	\label{fig:enhanced_signal_chain}
\end{figure}

Once the super-resolution FCNN has been trained by the proposed technique, our novel tracking algorithm can be implemented using the particle filter discussed previously.
The Doppler-corroborated particle filter is employed to track the position of the hand in the $y$-$z$ plane, and two additional particle filters are used to track the Doppler velocity and cross-range oscillation.
The entire signal processing chain for the enhanced feature extraction and tracking method is shown in Fig. \ref{fig:enhanced_signal_chain}.
The spatiotemporal features are outputted from the algorithm and can be used for many tracking applications.
Additionally, if the 2-D location of the hand is desired over the range and cross-range oscillation rate, the algorithm can be easily adapted to output the desired spatial features.

\section{Spatiotemporal Feature Extraction and Tracking Results}
\label{sec:results}

In this section, we overview the results of our novel tracking and feature enhancement algorithms beginning with the simple, classical techniques and compare the performance to our proposed methods.
Our enhanced tracking regime demonstrates considerable performance improvement compared with the traditional methods and allows for robust super-resolution tracking on a small radar platform unattainable by existing methods.

\subsection{Ground Truth - Ideal Motion Profile}
\label{subsec:ideal_motion_profile}

\begin{figure}[h]
	\centering
	\includegraphics[width=0.75\textwidth]{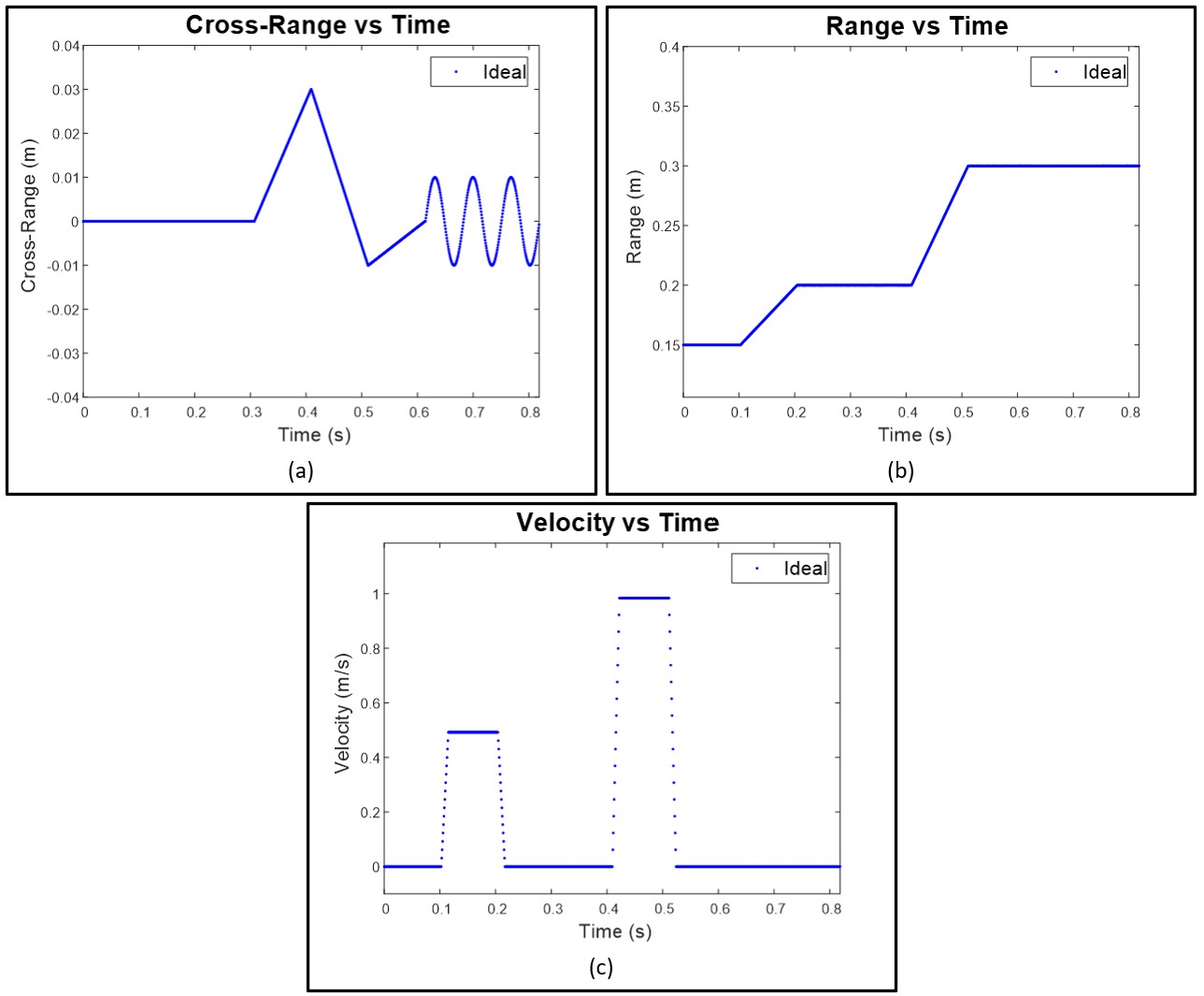}
	\caption{Ideal motion profile of the target in the (a) cross-range and (b) range directions as well as the (c) range velocity profile against time.}
	\label{fig:ideal_motion}
\end{figure}

To verify the feature estimation techniques, a virtual prototyping approach is adopted. A point target is simulated in motion with $y$-$z$ location and velocity shown in Fig. \ref{fig:ideal_motion} using (\ref{eq:mimo_beat_signal}). 
This ideal motion profile is employed to compare the tracking performance of our proposed methods to the traditional techniques. 
Real noise collected from the radar with an empty scene is added to each synthetic beat signal as
\begin{equation}
\label{eq:beat_sim_with_noise}
	\tilde{s}(y_T,y_R,k) = \frac{p}{R_T R_R}e^{jk(R_T + R_R)} + \alpha \tilde{ \omega}(y_T,y_R,k),
\end{equation}
where $\tilde{ \omega}$ is a complex-valued noise sample corrupting the amplitude and phase of the ideal simulated beat signal and $\alpha$ controls the SNR.

The motion profile shown in Fig. \ref{fig:ideal_motion} shows the ideal range ($z$), cross-range ($y$), and velocity ($v$) of the target. 
The motion profile includes independent and joint movement in the range and cross-range domains in addition to sinusoidal cross-range oscillation.
For our simulations, $4096$ time samples are generated using $p \in [0.5,1]$ to simulate the variance in the hand's empirical radar cross-section (RCS) as observed from prior hand data and $\alpha \in [1,3]$ to vary the SNR among samples. 
Values for $p$ and $\alpha$ are selected randomly within the specified intervals for each time sample and provide a level of stochastic realism to the simulated data.

\subsection{Classical Spatiotemporal Imaging Results}
\label{subsec:simple_gesture_tracking_results}
First, the simple tracking methods discussed in Section \ref{subsec:classical_gesture_tracking} are implemented to provide baseline performance metrics.
The signal processing chain shown in Fig. \ref{fig:simple_signal_chain} is performed, extracting the spatiotemporal features. 
At each iteration, the features are extracted directly from the raw RMA images and are therefore prone to erratic behavior. 

\begin{figure}[h]
	\centering
	\includegraphics[width=0.75\textwidth]{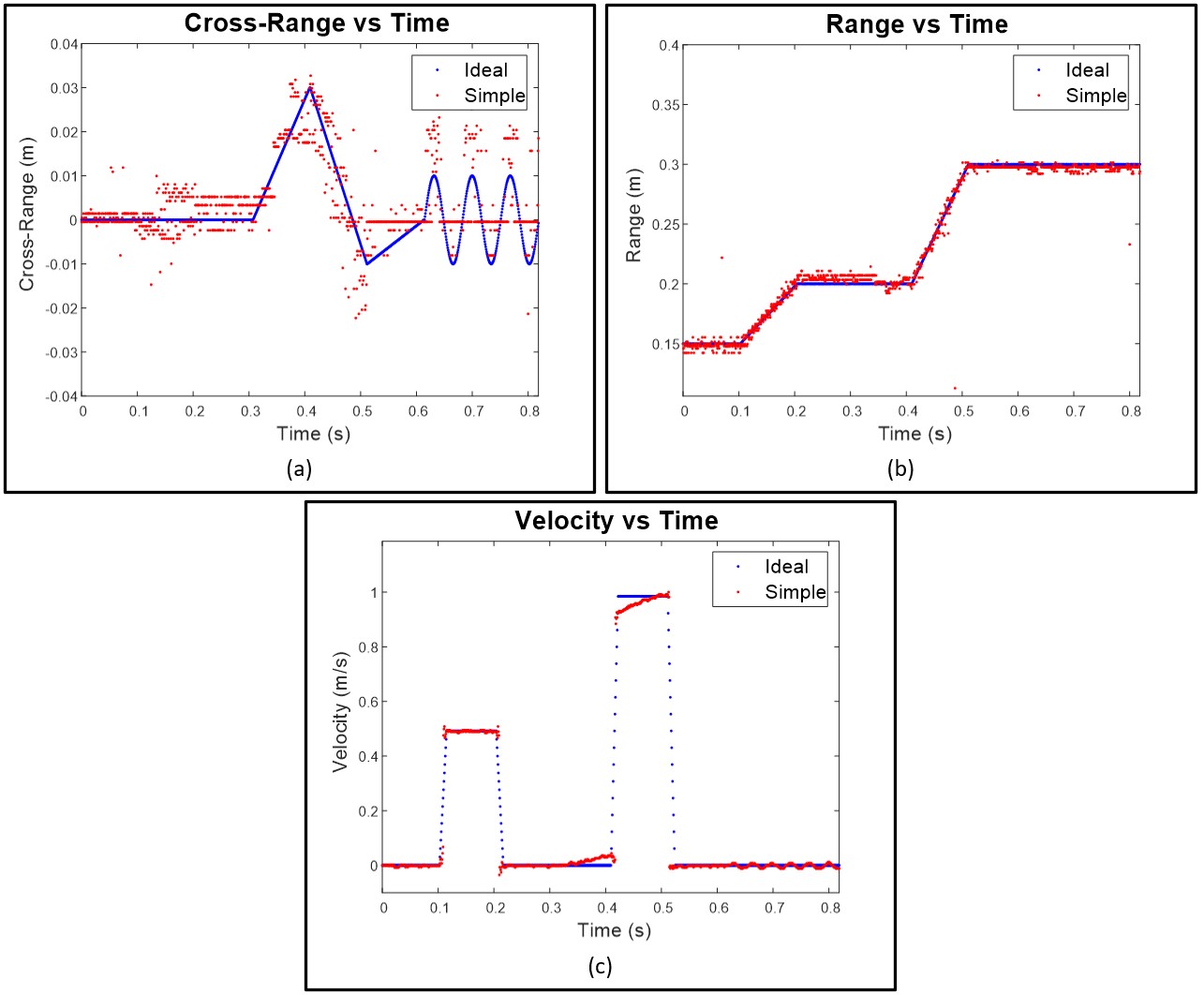}
	\caption{Motion profile using simple feature extraction techniques on each frame for every time step (red) compared with the ideal motion and velocity profiles (blue). The (a) cross-range and (b) range are measured directly from the peak of the RMA image of each frame and the (c) velocity is measured using the Doppler FFT of the raw RMA images using (\ref{eq:doppler_fft}) and (\ref{eq:velocity_doppler}).}
	\label{fig:simple_motion}
\end{figure}

Fig. \ref{fig:simple_motion} shows the features estimated from the data generated by (\ref{eq:beat_sim_with_noise}) using the simple methods. 
The real radar noise and varying reflectivity result in outliers and errors in the estimated location and velocity of the target, particularly in the cross-range domain.
Without more robust feature extraction and tracking techniques, the performance leaves much to be desired.
In the following sections, the performance of the simple tracking methods is quantitatively compared to the enhanced tracking methods and design considerations are discussed.

\subsection{FCNN-Based Super-Resolution Tracking Results}
\label{subsec:enhanced_gesture_tracking_results}
Assuming the motion profile in Fig. \ref{fig:ideal_motion}, our proposed particle filter algorithm is employed in an attempt to more robustly track the 2-D position and Doppler velocity of the target across time, improving the user's control over the interface significantly.

\begin{figure}[h]
	\centering
	\includegraphics[width=0.75\textwidth]{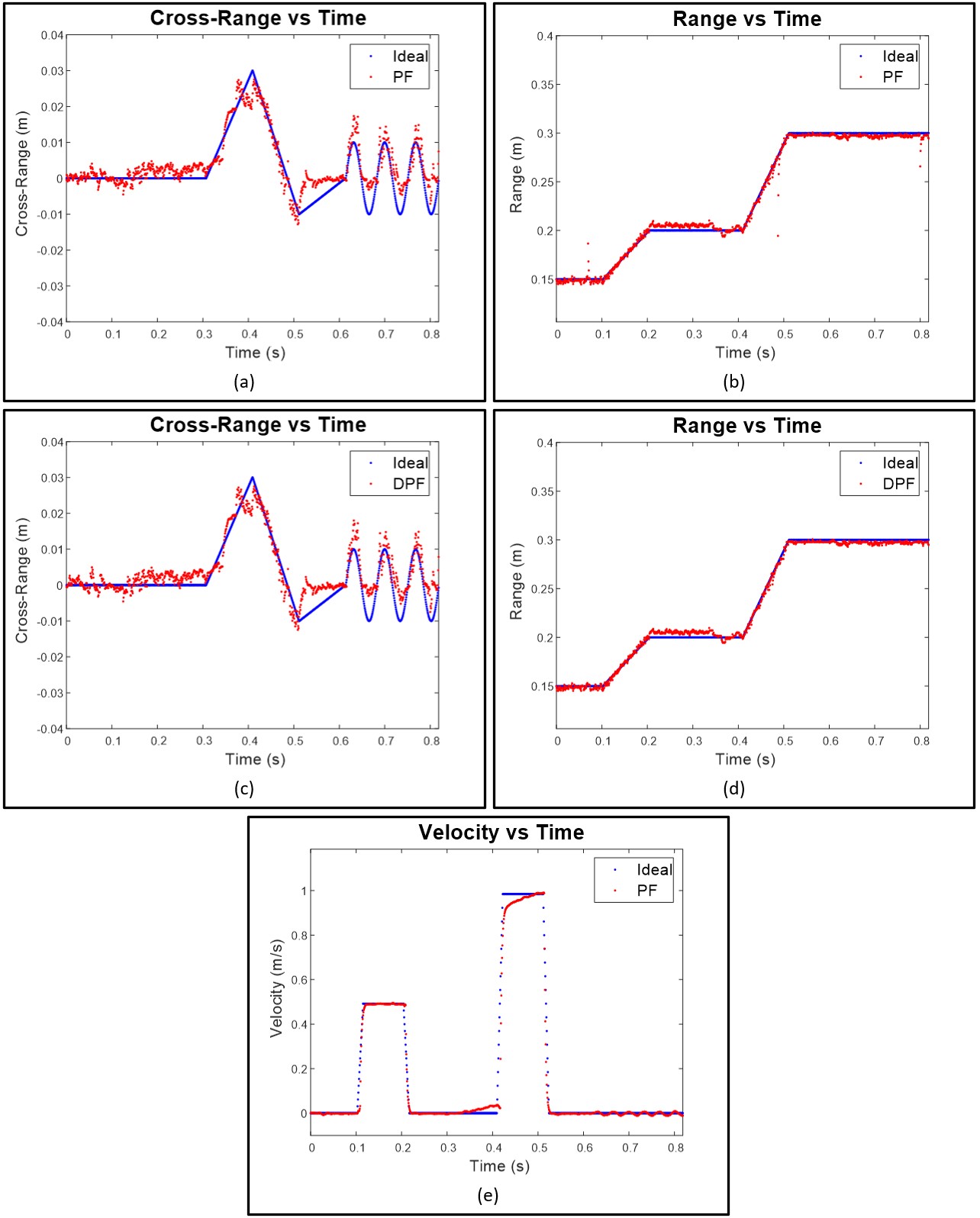}
	\caption{Particle filter (PF) and Doppler-corroborated particle filter (DPF) tracking. Improved tracking of the (a)/(c) cross-range and (b)/(d) range positions versus time using the PF/DPF with $N_z =$ 16, and (e) Doppler velocity versus time using a PF approach.}
	\label{fig:pf_motion}
\end{figure}

First, the particle filter algorithm (PF) without Doppler corroboration is implemented using the data in Fig. \ref{fig:simple_motion} as elements of the noisy measurement vector $\mathbf{r}$. 
The PF reduces the effect of the noise on the position estimation and improves the spatiotemporal tracking performance as shown in Fig. \ref{fig:pf_motion}.
The cross-range position tracking is most improved compared to the traditional methods.
Next, the Doppler-corroborated particle filter (DPF) is applied to the same set of data further improving the estimation of the range. 
The outliers in Fig. \ref{fig:pf_motion}b are mitigated by the DPF in Fig. \ref{fig:pf_motion}d because the outlying samples result in a sample velocity $\hat{v}_s$ contradicted by the Doppler velocity $\hat{v}_d$ and are weighted as unimportant in the resampling process.
The DPF algorithm improves the user experience of our interface by providing a robust, consistent tracking algorithm to smoothly estimate the 2-D position and spatiotemporal signatures of the user's hand.
However, the PF and DPF can be further improved by implementing the proposed enhancement FCNN.

\begin{figure}[h]
	\centering
	\includegraphics[width=0.85\textwidth]{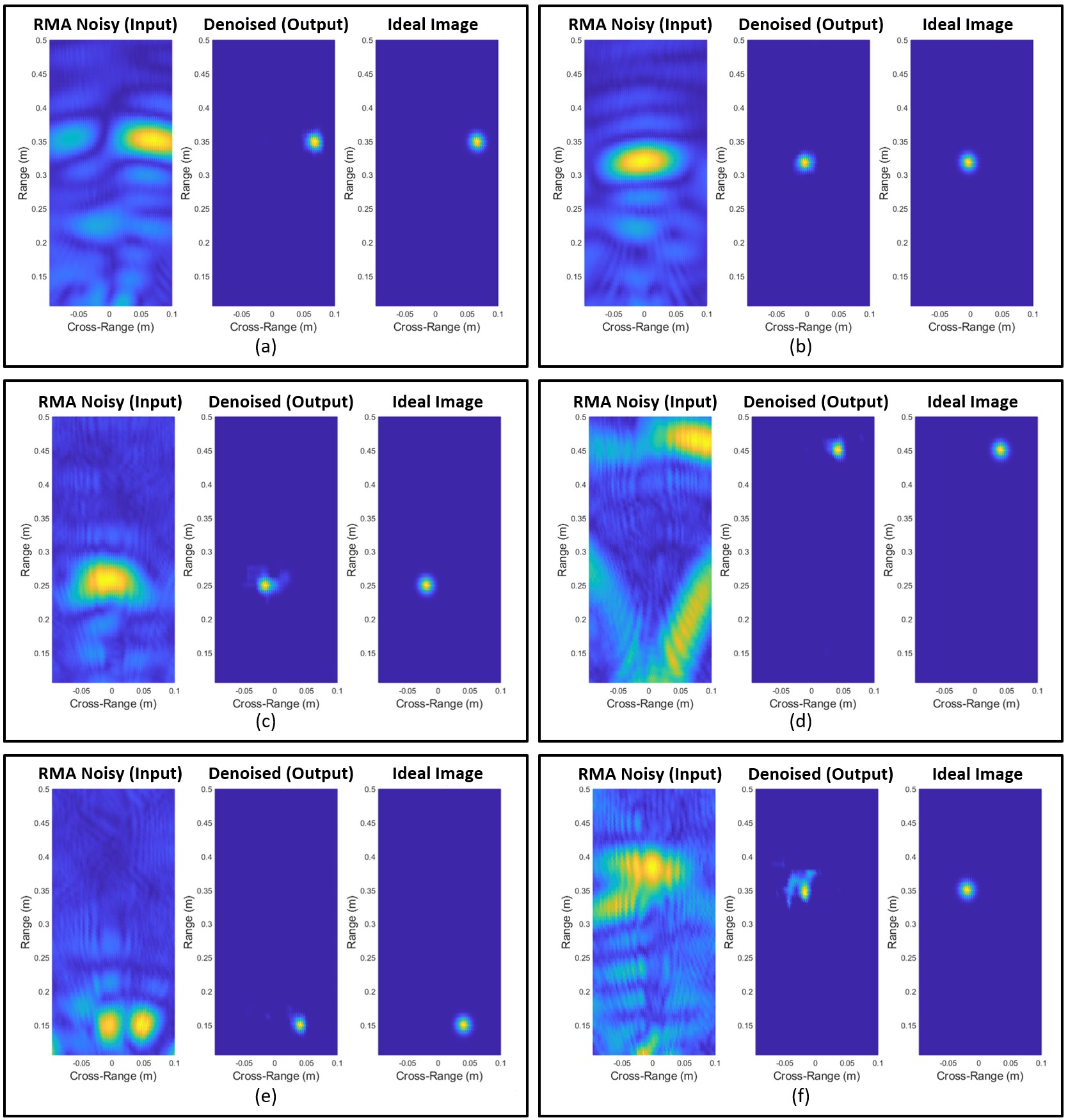}
	\caption{Enhancement FCNN applied to simulated (a,b) and real hand (c-f) RMA images for image enhancement and improved localization.}
	\label{fig:fcnn_enhancement_demo}
\end{figure}

After the super-resolution FCNN is trained using the technique discussed in Section \ref{subsec:enhanced_tracking_implementation}, a validation dataset of identical size to the training set is collected.
Fig. \ref{fig:fcnn_enhancement_demo} shows images enhanced by the enhancement FCNN demonstrating the robustness of the network. 
Figs. \ref{fig:fcnn_enhancement_demo}a and \ref{fig:fcnn_enhancement_demo}b show simulated point targets enhanced by the FCNN resulting in localization super-resolution. 
Fig. \ref{fig:fcnn_enhancement_demo}c is an RMA image reconstructed from a real hand capture close to the middle of the cross-range domain. 
The 2-D position of the hand is accurately located compared with the ideal image. 
Similarly, Figs. \ref{fig:fcnn_enhancement_demo}d-\ref{fig:fcnn_enhancement_demo}f demonstrate the network's ability to enhance images degraded by small hand RCS in comparison to noise, ghosting due to non-ideal beam patterns, ambient and device noise, and other non-idealities.
The proposed enhancement FCNN simultaneously enables localization super-resolution and overcomes device and environment issues.
Hence, the features extracted from the enhanced images are much improved compared to the raw RMA images before the FCNN and result in superior tracking performance.

\begin{table} [h]
	\caption{Simple vs Enhanced Localization RMSE}
	\centering
	\begin{tabular}{| c || c |  c |}
		\hline
		& $y$ (m) & $z$ (m) \\
		\hline\hline
		Simple & 0.0154 & 0.023 \\ 
		\hline
		Enhanced & 0.0085 & 0.0083 \\ 
		\hline
	\end{tabular}
	\label{table:fcnn_position_rmse}
\end{table}

To quantitatively compare the localization improvement of the enhancement FCNN compared to the simple method, the RMSE in the range and cross-range position are computed on the validation dataset using the two techniques and shown in Table \ref{table:fcnn_position_rmse}. 
The enhancement FCNN improves both the resolution of the RMA images and the localization accuracy for both simulated and real data. 

\begin{figure}[h]
	\centering
	\includegraphics[width=0.75\textwidth]{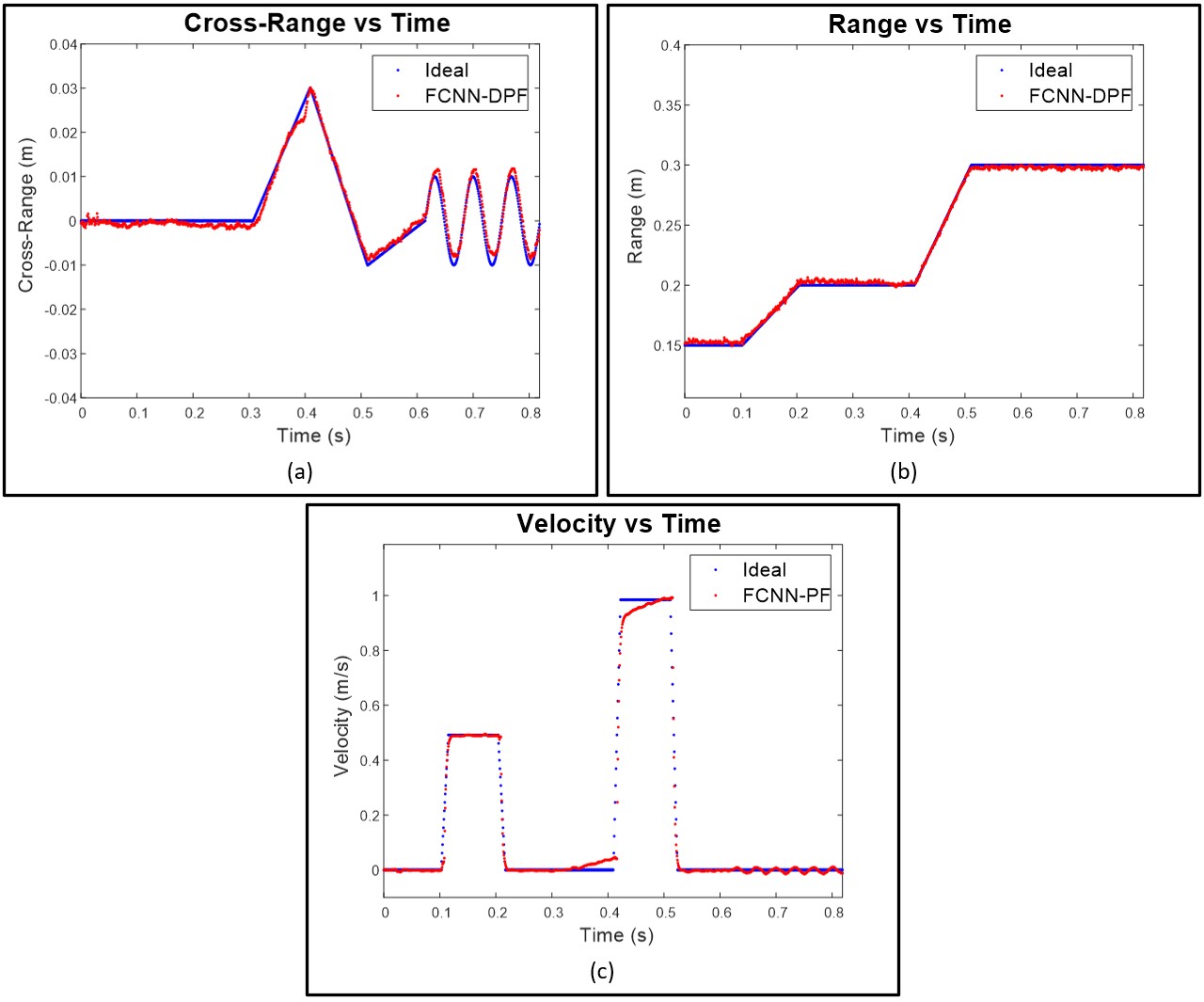}
	\caption{Spatiotemporal tracking with the FCNN-enhanced Doppler-corroborated modified particle filter algorithm.}
	\label{fig:fcnn_dpf_motion}
\end{figure}

Fig. \ref{fig:fcnn_dpf_motion} demonstrates the tracking performance of the FCNN and DPF (FCNN-DPF) on the same data as the previous tracking examples, an noted improvement over the DPF alone. 
Applying the FCNN-DPF, the range and cross-range tracking of the target is nearly identical to the ideal motion profile and an improvement in the velocity estimation.
Using the identical sporadic data resulting in the poorly estimated cross-range positions in Fig. \ref{fig:simple_motion}a, the FCNN-DPF yields an estimation nearly identical to the ideal motion profile.
Similarly, the cross-range estimates in Fig. \ref{fig:pf_motion}a and Fig. \ref{fig:pf_motion}c are outperformed by the FCNN-DPF in Fig. \ref{fig:fcnn_dpf_motion}a.
Compared to the classical techniques and PF/DPF alone, the localization performance of the FCNN-DPF is considerably superior.

\begin{figure}[h]
	\centering
	\includegraphics[width=0.5\textwidth]{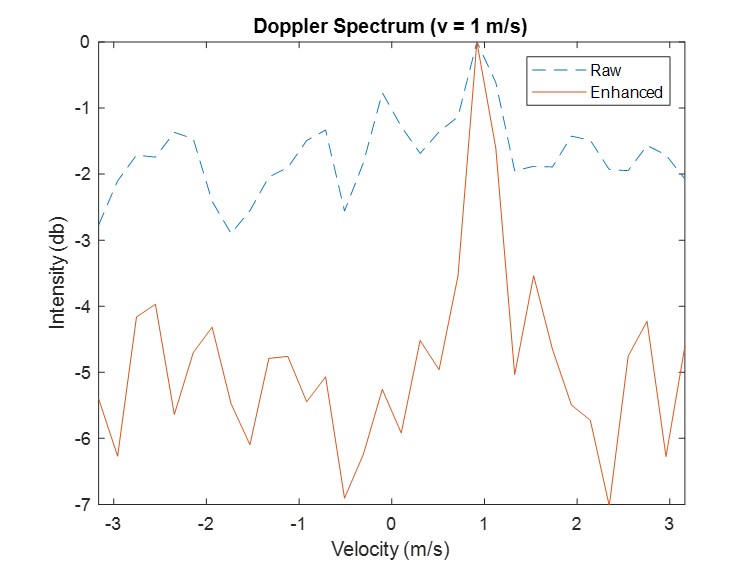}
	\caption{Comparison of the Doppler velocity spectrum when the Doppler FFT and video pulse integration steps are performed on the raw RMA images compared to the enhanced RMA images. The simulated data contains 128 frames and uses $\alpha = $ 3 for every capture to simulate a low SNR scenario.}
	\label{fig:doppler_snr}
\end{figure}

Further, the FCNN improves the Doppler estimation robustness. 
As shown in Fig. \ref{fig:doppler_snr}, the Doppler spectrum SNR is improved when the Doppler processing is performed on the enhanced RMA images as compared to Doppler processing on the raw RMA images.
Hence, the enhancement network improves the reliability of the Doppler velocity estimation aiding spatiotemporal tracking.

\section{Discussion and Future Work}
\label{sec:discussion}

To quantitatively compare the tracking performance of the various proposed methods, $4096$ unique motion profiles are generated and corresponding tracking RMSE is computed for the cross-range, range, and velocity. 
Displayed in Table \ref{table:tracking_rmse2}, the RMSE for the cross-range ($y$), range ($z$), and velocity ($v$) improve with the novel algorithms proposed in this article. 

As expected, the baseline simple method yields the greatest error for all three features. 
Comparing PF and DPF, the cross-range and velocity RMSE are identical between the two techniques but the range RMSE is improved due to the dynamic weighting technique. 
The FCNN alone outperforms the simple method but can be improved by including the PF and DPF after image enhancement. 
Finally, the FCNN-PF and FCNN-DPF yield identical results for the cross-range and velocity RMSE, as expected, but significant improvement can be noted in the range error. 
The results in Table \ref{table:tracking_rmse2} demonstrate the considerably superior tracking performance of the enhanced tracking methods, namely the FCNN-DPF, compared with the simple tracking method.
The performance gain realized by implementing the super-resolution FCNN demonstrates the ability of the network to learn the system noise and ambiguities during the training phase using both real and synthetic data. 

The average latency of each method, $\Bar{\tau}$, is measured as the time duration between the new sample being captured and the estimation process being completed on that sample. 
The resulting estimates are streamed across the MIDI port or sent to the built-in audio signal generation tool. 
Additional latency contributed by the subsequent synthesis engine is highly dependent on the software used and device under test; thus, it is not considered as part of the latency due to our methods.

\begin{table} [h]
	\caption{Average RMSE for Tracking Methods}
	\centering
	\begin{tabular}{ | c || c | c | c | c | }
		\hline
		& $y$ (mm) & $z$ (mm) & $v$ (mm/s) & $\Bar{\tau}$ (ms) \\
		\hline\hline
		Simple & 7.86 & 22.0 & 72.4 & 2.29 \\
		\hline
		PF & 5.27 & 13.6 & 52.9 & 2.36 \\
		\hline
		DPF & 5.27 & 6.85 & 52.9 & 2.41 \\
		\hline
		FCNN & 7.74 & 12.3 & 58.4 & 2.67 \\
		\hline
		FCNN-PF & 3.70 & 7.44 & 44.5 & 3.92 \\
		\hline
		FCNN-DPF & 3.70 & 3.07 & 44.5 & 3.96 \\
		\hline
	\end{tabular}
	\label{table:tracking_rmse2}
\end{table}

The enhanced tracking methods outperform the existing techniques in localization resolution, Doppler spectrum SNR, and tracking accuracy; however, there are some necessary trade-offs for this performance gain. 
The novel super-resolution FCNN yields noteworthy resolution improvement over the theoretical bounds.
In ideal conditions, the cross-range and range resolutions of our system are bounded by $\delta_y = 7.5$ cm and $\delta_z = 3.75$ cm, respectively.
Using the combination of real hand data and synthetic data from Table \ref{table:fcnn_position_rmse}, the spatial resolution in each direction is computed empirically as $\delta_y = 2.3$ mm and $\delta_z = 1.96$ mm. 
On the other hand, the effectiveness of the enhancement FCNN is limited by the training set. 
Since the FCNN is only trained on images within the expected region, extending the ROI outside of the trained region results in performance degradation. 
If the ROI is changed, the FCNN should be retrained accordingly.
In contrast, the simple methods are highly flexible but cannot compete with the performance of the enhancement techniques.
However, we have studied the limitations of the particular TI mmWave radar device and found that if the hand is placed outside the ROI defined in the previous section, it will not be detected.
Due to device SNR and beamwidth, for most hand sizes, the reflections back to the radar will not be strong enough for detection.
Additionally, we have tested the proposed FCNN in smaller ROIs and found similar results without retraining.
For other array topologies, the proposed methods can be easily applied, although the FCNN will need to be trained accordingly.

While the Doppler-corroborated particle filter improves the tracking robustness, it requires a high throughput framework to function properly.
Since the DPF relies on accurate Doppler velocity estimation, the pulse repetition interval PRI, $T_{PRI}$, must be sufficiently small such that high hand velocities are within the resolvable range.
The PRI is impacted most significantly by the time per iteration of the signal processing chain.
Given our framework is currently released in the prototyping stage as a MATLAB program, the latency performance does not match that of a real-time implementation on a more efficient embedded device.
Hence, typical throughput times limit the PRI to around $4$ ms.
At $T_{PRI} = 4$ ms using a $77$ GHz device, the maximum resolvable velocity is $0.24$ m/s.
With this limitation, some rapid movements at high velocities may result in Doppler spectrum aliasing.

While the software package presented in this article serves as a framework for demonstrating and prototyping the proposed tracking and super-resolution algorithms, the inherent latency of the signal processing steps is a key issue in HCI and must be addressed.
In our research, the largest contributor of latency in our proposed system is the hand-off between the radar device and MATLAB over UDP and shared memory, at an average of $1.93$ ms. 
Rather than streaming to data MATLAB, a real-time solution can be implemented on the TI radar device's built-in DSP, thus providing a more efficient throughput as the DSP has direct access to the samples as they are taken. 
Additionally, several steps in the signal processing chain will increase in efficiency with an embedded solution.
Employing small window sizes, $N_z = 16$ and the number of FFT spatial points is $64$, the DPF and FFT computation times can be further reduced compared to the relatively inefficient MATLAB implementation.
We would also like to note that a significant decrease in latency was achieved by optimizing the implementation using GPU accelerated coding.
A similar approach could be taken on an embedded solution leveraging the highly parallelizable nature of many of the steps in the signal processing chain (FFT, CNN, Gaussian distribution computation).
Comparing the computational efficiency among the algorithms, the latency cost for the more robust algorithms is insignificant in proportion to the performance gain, even in the MATLAB implementation.
In latency tests, the average response time using the FCNN-DPF was $3.96$ ms from user input to MIDI signaling.
While most MIDI interfaces outperform this metric, we believe our framework demonstrates a competitive throughput cycle time compared to existing technology and can be further improved by a more efficient implementation.

Hand-tracking using a mmWave radar has both advantages and drawbacks compared with other sensing regimes.
In this article, we employ a single radar to develop and demonstrate robust tracking algorithms for mmWave devices.
While the best performance is likely achieved through a sensor fusion technique, a radar-based implementation may be optimal if privacy is a concern using optical cameras or issues such as occlusion and lighting conditions must be taken into account.
Compared to optical and RGB+D solutions, mmWave is more versatile and reliable, operating well under occlusion, in any temperature or lighting environment, and offers precise depth information of the entire scene.
For a musical interface, these advantages may not be often fully realized; however, the novel tracking methods proposed in this article are applicable for many HCI applications.
On the other hand, mmWave sensors cannot meet the performance of optical solutions when it comes to cross-range resolution due to the limited aperture size, making multi-object and finger tracking much more challenging. 
As such, many applications in HCI, computer vision, automated driving, etc. employ radar, lidar, and optical imaging devices with sensor fusion algorithms to achieve further improved performance at an increased cost. 
For these applications, our proposed algorithms can aid in sensor fusion by significantly increasing the performance contribution from the radar sensors.

Several alternatives exist to mmWave radar sensing, namely wearable, handheld, and optical devices. 
Wearable and handheld sensing solutions offer highly precise spatiotemporal features but are often not preferable compared to contactless sensors \cite{pardue2013hand,neto2010high}.
In terms of cost, mmWave radar devices are in the same price bracket as the popular Kinect and Leap Motion optical sensors on the order of $\$100-\$200$.
Attempts using multiple RGB cameras \cite{ballan2012motion,sridhar2013interactive} show promising results; however, a single device is much preferred due to the cumbersome nature of multi-camera systems. 
Single RGB+D solutions have been proposed using generative pose tracking \cite{oikonomidis2011efficient,tang2015opening} and learning-based generative pose tracking \cite{sridhar2015fast,taylor2016efficient}.
However, all of these methods suffer tremendously under occlusion or scene clutter, both of which can be overcome using mmWave radar. 
Some deep learning-oriented solutions have shown quite promising results \cite{tompson2014real,ye2016spatial}, but constructing a sufficient dataset for meaningful supervised training remains a challenge. 

Our proposed interface tracks the 2-D position and velocity of the user's hand to control note selection and two user-selected parameters, a marked improvement over the prior work on mmWave radar using the Google Soli tracking only 1-D range for parameter control \cite{bernardo2017_o_soli_mio}. 
However, optical solutions enable tracking of both hands \cite{polfreman2011multi,jensenius2013kinectofon,hantrakul2014implementations,tompson2014real,ye2016spatial,sridhar2015fast,taylor2016efficient,oikonomidis2011efficient,tang2015opening} or hand and finger position \cite{han2014lessons,nieto2013hand} for even finer musical control, with some scenario-specific drawbacks. 
As radar technology improves and larger apertures become widely available, tracking individual fingers will become increasingly plausible and could yield comparable or superior results to optical solutions due to superior depth resolution.

Compared to prior work on hand-tracking with mmWave devices, our proposed methods yield competitive results.
Past work using radar devices achieves, at best, an average range tracking error of $2$ cm on human hand localization \cite{joshi2015wideo}. 
Our enhanced tracking technique yields a mean range tracking error of $1.89$ mm, improving tracking by more than a factor of ten.
In \cite{li2020thumouse}, a $4$ GHz bandwidth mmWave sensor achieves a 2-D position RMSE of $1.16$ mm tracking a thumb, at distances closer than $10$ cm.
Comparatively, our enhanced tracking technique tracks a human hand across much larger distances and still achieves a competitive 2-D position RMSE of $3.4$ mm. 
At the time of this article, we are not aware of any other prior work on hand-tracking using mmWave devices. 
To our knowledge, the system proposed in this article offers unprecedented hand gesture tracking performance using a single mmWave sensor.

The most direct musical interface comparison to our framework is the Theremin, as both are controlled by the hand's proximity to the sensor. 
The pitch of the Theremin is controlled continuously by the hand's vertical location, whereas our interface tracks the range of the hand digitally and selects a note from the user-defined scale. 
While the Theremin uses two antennas, one for volume control and the other for pitch control, a total of two degrees of freedom, our framework offers three degrees of freedom (range, cross-range, and velocity), thus providing three controllable parameters.
As previously mentioned, the musical interface promoted in this article supports Theremin-like gestures for note selection and parameter control.
However, high-velocity percussive gestures could be implemented using our high-fidelity tracking algorithms, with some limitations.
Small values of the weighting vector, $\mathbf{a}$, in the particle filter algorithm can result in an excessively smoothed and overly damped system limiting the ability of the system to track sudden movements. 
Depending on the desired application, finely tuning this parameter is essential for enabling proper gestural control.
Our proposed interface is an evolved Theremin, utilizing a modern mmWave sensor for precise tracking in 2-D space (expansion to 3-D can be easily implemented with the proper hardware). 
In contrast to a Theremin, our musical interface is significantly less effortful in note selection, allowing simple and intuitive inclusion of the additional parameter controls and increasing accessibility to the user base.
One of the authors is a skilled guitar and violin instrumentalist with a background in electronic music production.
From the perspective of an experienced musician, the proposed methods provide the musician a sufficient and consistent level of control and offer an elegant new musical interface capable of generating unique phrases previously only possible by transcribing MIDI notes and control parameters manually into a digital audio workstation (DAW) or another live digital synthesis platform. 

For future work, several promising routes are left to be explored. 
First, further development can be explored by implementing our proposed methods onto a real-time embedded platform.
Additionally, using multiple MIMO radars or a larger MIMO array, a multiple-hand and individual finger tracking interface can be investigated, thus further extending the application space of our robust tracking methods.
Finally, our novel super-resolution tracking algorithms can easily be adapted to offer an elegant, efficient solution to a host of acute hand-tracking problems in the HCI domain and even employed in sensor-fusion systems. 

\section{Conclusion}
\label{sec:conclusion}

Our FCNN-based super-resolution framework successfully demonstrates the viability of acute human hand-tracking for HCI using mmWave sensors. 
We validated and implemented our spatiotemporal signal processing algorithms and robust tracking algorithms in the form of a contactless musical interface; however, this article also serves to demonstrate the broad effectiveness of mmWave technology for a multitude of near-field acute hand-tracking applications. 
First, simple feature extraction and tracking methods were introduced, followed by an enhanced approach leveraging the Doppler-corroborated particle filter algorithm and enhancement FCNN to achieve robust tracking and super-resolution in a non-ideal imaging scenario.
The methods are compared demonstrating noticeable improvement using the FCNN-DPF over the classical techniques.
Additionally, our work offers competitive tracking estimation and localization performance compared to prior methods in the literature for both mmWave and optical implementations. 
Our entire software implementation and real-time radar interface platform are freely available at request.
The novel FCNN-based super-resolution and tracking algorithms presented in this article offer an elegant solution to many contactless HCI problems.

\section*{Acknowledgment}
The Josiah Smith's work was supported by the imec USA summer internship program. The authors want to thank Dr. Gonzalo Vaca Castano for his insights in developing the particle filter algorithm and computer vision approach. The work of Murat Torlak (while serving at NSF) was supported by the NSF. Any opinions, findings, and conclusions or recommendations expressed in this material are those of the author(s) and do not necessarily reflect the views of the NSF.

\bibliography{ref_list}
\bibliographystyle{IEEEtran}

\end{document}